\documentclass{aa}
\usepackage{graphicx}
\usepackage{txfonts}
\usepackage{natbib}
\bibpunct{(}{)}{;}{a}{}{,}  
\def\etal   {{\it et~al.$\;$}}
\def\ie {$i.e.,\;$}
\def\eg {$e.g.,\;$}
\def\ie {$i.e.,\;$}
\def\eg {$e.g.,\;$}
\def\viz {$viz.,\;$}

\def\abt {$\sim$}
\def\deg {$^\circ$}

\def\La1215 {Ly$\alpha\lambda1215$}

\begin{document}
   \title{Milliarcsec-scale radio structure of a matched sample of Seyfert~1
 and Seyfert~2 galaxies\thanks{The other 24 panels of Fig.~1 are only
 available in electronic form. The full paper copies can be obtained from
 the authors.}{$^,$}\thanks{Tables 3--4 are only available in electronic form
 and Tables 5--6 are also available at the CDS via anonymous ftp to
 cdsarc.u-strasbg.fr {\emph{(130.79.128.5)}} or via
 http://cdsweb.u-strasbg.fr/cgi-bin/qcat?J/A+A/}}


   \author{D. V. Lal \inst{1,2}{$^,$}\thanks{\emph{Present address:}
          National Centre for Radio Astrophysics, Post Bag 3, Ganeshkhind,
          Pune 411 007, India}
          P. Shastri \inst{1},
          \and
          D. C. Gabuzda \inst{3}
          }

   \offprints{Dharam Vir Lal}

\institute{Indian Institute of Astrophysics, Koramangala,
              Bangalore -- 560 034, India\\
              \email{dharam@ncra.tifr.res.in}
         \and
             Department of Physics, Indian Institute of Science,
              Bangalore -- 560 012, India
         \and
             Department of Physics, University College Cork, Cork, Republic
             of Ireland
             }

   \date{Received ; accepted }

   \abstract{
We have obtained mas-scale resolution very long baseline interferometry
(VLBI) images of a sample of Seyfert~1 and Seyfert~2 galaxies at 5~GHz
(wavelength, $\lambda$~=~6~cm). The Seyferts
of the two types were chosen to be matched in several orientation-independent
properties, primarily in order to rigorously test predictions of the unified
scheme. We detected all the 15 objects that we observed. In this paper we
describe the observations and data reduction procedures, and present the VLBI
radio images as well as simultaneous Very Large Array images that we
obtained for these 15 Seyferts.
   }

\authorrunning{Lal, D.V. \etal}
 \titlerunning{Milliarcsec-scale radio structure of Seyferts}
   \maketitle

\section{Introduction}
\label{intro}

Seyfert galaxies have been long-recognized as a class of active galactic
nuclei (AGN) that are at the low end of the luminosity range (and the distance
range) of AGN, and primarily characterized by optical emission lines that
have a wide range of ionization \citep[\eg][]{Osterbrock89}.
Usually the term ``Seyfert galaxy" is used to refer to radio-quiet AGN that
are in spiral host galaxies.
The current majority-view is that the nuclei of Seyfert galaxies are powered
by super-massive black holes, although
there is also the alternate view that star bursts {\it alone} can
power Seyferts (\eg \citet*{CidFernandes97} and refs therein).

\cite{KhachikianWeedman74} identified two types of Seyfert
galaxies on the basis of the widths of the nuclear emission lines.
While spectra of type~2 Seyfert galaxies have relatively narrow emission lines
(FWHM~$\approx$~300~km~s$^{-1}$ to 1000~km~s$^{-1}$),
the hydrogen and helium lines in the
spectra of type~1 Seyfert galaxies have an additional much
broader component (FWHM~$\ge$~1000~km~s$^{-1}$).
The Unified Scheme (US) for Seyfert galaxies
\citep[\eg][]{Antonucci93} hypothesizes that Seyfert~1s and 2s constitute
the same parent population and appear different {\it solely} due to their
differing orientations. The US requires the presence of optically thick
material around the central region in the form of a torus
\citep[\eg][]{PierKrolik93II}.  When the torus is face-on, the
broad emission lines from the central clouds are directly
visible, as in Seyfert~1s, whereas when the torus is edge-on, it obscures these
clouds from view, resulting in Seyfert~2s. The US  found particularly strong
support in the discovery of {\it polarized} broad emission lines from Seyfert~2s
(\citet{AntonucciMiller85,Moran00}), which are interpreted as
originating from the hidden clouds but scattered  periscopically into our
line of sight.

The US has definite predictions for the radio morphology of
Seyferts: the total radio luminosities of Seyfert~1s and 2s are expected to be
similar and the radio structures are expected to differ only by
projection effects. Indeed, Seyfert~1s and 2s have been shown to have
similar luminosities on various spatial scales \citep{UlvestadWilson89}.
The result of \cite{Roy94}, however, contradicts
the US. They reported that Seyfert~2s, \ie those that have
edge-on tori in the US, have a larger likelihood of showing compact
radio emission than Seyfert~1s. On the other hand, the simple US predicts
that there should be no such difference between the Seyfert subclasses.

The intriguing result of \citeauthor{Roy94} was obtained using relatively large
samples of Seyferts, but the compact structure was measured using a
275~km interferometer between the Parkes and Tidbinbilla telescopes.
We sought to improve upon their experiment by using pc-scale, \ie Very Long
Baseline Interferometry (VLBI) {\it images} of Seyferts rather than a
single-baseline measurement of the visibility and by designing our sample
so as to enable {\it rigorous} tests of the US.
Our sample selection is described in detail in an accompanying
paper. The key point is that the Seyfert~1s and Seyfert~2s were
chosen to be {\it matched} in a range of orientation-independent
properties, so that we can compare objects of the two types that should be
{\it intrinsically similar} in the framework of the US.

Our sample has 10 Seyfert~1s and 10 Seyfert~2s, including 15 objects that
did not have prior VLBI observations. We present here images
derived from global VLBI and simultaneous Very Large Array
(VLA) observations of these 15 objects at 5~GHz.
In a later paper, when interpreting our results, we will combine
our data with previously published data for the remaining
five objects of our sample (see also \cite{Shastri03I}).

\section{Observations}
\label{observe}

We obtained global VLBI observations of the 15 Seyfert galaxies
(see Table~\ref{source_info})
in our sample that did not have prior VLBI observations published
in the literature during a single 24~hour run on February 18, 1998.
Since the phased VLA was one of our VLBI stations, we also
obtained simultaneous VLA aperture-synthesis data for 
all these 15 sources.

\subsection{VLBI observations}
\label{gl022_obs}

The global VLBI array consisted of a total of 14 stations:
three stations of the European VLBI Network (EVN), \viz Effelsberg
(Germany), Torun (Poland) and Noto (Italy), all 10 stations of the Very
Long Baseline Array (VLBA) and the phased VLA. The inclusion of
two stations with relatively large collecting area, \viz Efflesberg and
the phased VLA, was critical to our ability to detect the
radio-faint Seyferts in our sample. Information about the global
VLBI observations is summarized in Table~\ref {vlbi_log}.

\begin{table*}[]
\caption{Seyfert galaxies that were observed using global VLBI.
References for the optical positions:
$a$: \cite{Clements81}; $b$: \cite{Clements83}; $c$: \cite{ArgyleEldridge90};
$d$: \cite{Gallouet75}.
{We use H$_0$ = 75 km~s$^{-1}$ Mpc$^{-1}$ and q$_0$ = 0}.}
\begin{tabular}{l|cc|c|c|c}
\hline
 & \multicolumn{2}{|c|}{Position of optical nucleus} & & & \\ \cline{2-3}
 Object & \multicolumn{2}{|c|} {Right Ascension  ~~~Declination} & Redshift & Distance & Image scale \\
 &\multicolumn{2}{c|}{(J2000)}& & (Mpc) & (pc mas$^{-1}$) \\ \hline
&&&&& \\
Mrk 1      &01$^h$16$^m$07$^{s}$.25 & $+$33$^\circ$05$^\prime$22$^{\prime\prime}$.40$^a$&0.016&63.5& 0.31 \\
MCG 8-11-11&05$^h$54$^m$53$^{s}$.61 & $+$46$^\circ$26$^\prime$21$^{\prime\prime}$.61$^a$&0.025&98.8&0.48 \\
NGC 2273   &06$^h$50$^m$08$^{s}$.72 & $+$60$^\circ$50$^\prime$45$^{\prime\prime}$.01$^c$&0.006&23.9&0.12 \\
Mrk 78     &07$^h$42$^m$41$^{s}$.73 & $+$65$^\circ$10$^\prime$37$^{\prime\prime}$.46$^a$&0.037&145.4&0.69 \\
Mrk 1218   &08$^h$38$^m$10$^{s}$.95 & $+$24$^\circ$53$^\prime$42$^{\prime\prime}$.92$^b$&0.029&114.4&0.55 \\
NGC 2639   &08$^h$43$^m$38$^{s}$.00 & $+$50$^\circ$12$^\prime$20$^{\prime\prime}$.32$^c$&0.011&43.8&0.21 \\
Mrk 766    &12$^h$18$^m$26$^{s}$.51 & $+$29$^\circ$48$^\prime$46$^{\prime\prime}$.34$^a$&0.013&51.7&0.25 \\
Mrk 477    &14$^h$40$^m$38$^{s}$.11 & $+$53$^\circ$30$^\prime$16$^{\prime\prime}$.00$^a$&0.038&149.2&0.73 \\
NGC 5929   &15$^h$26$^m$06$^{s}$.13 & $+$41$^\circ$40$^\prime$14$^{\prime\prime}$.72$^b$&0.009&35.8&0.18 \\
NGC 7212   &22$^h$07$^m$02$^{s}$.01 & $+$10$^\circ$14$^\prime$00$^{\prime\prime}$.34$^b$&0.027&106.6&0.51 \\
Ark 564    &22$^h$42$^m$39$^{s}$.34 & $+$29$^\circ$43$^\prime$31$^{\prime\prime}$.31$^a$&0.024&94.9&0.46 \\
NGC 7469   &23$^h$03$^m$15$^{s}$.62 & $+$08$^\circ$52$^\prime$26$^{\prime\prime}$.39$^a$&0.016&63.5&0.31 \\
Mrk 530    &23$^h$18$^m$56$^{s}$.62 & $+$00$^\circ$14$^\prime$38$^{\prime\prime}$.23$^a$&0.030&118.3&0.57 \\
Mrk 533    &23$^h$27$^m$56$^{s}$.72 & $+$08$^\circ$46$^\prime$44$^{\prime\prime}$.53$^b$&0.029&114.4&0.55 \\
NGC 7682   &23$^h$29$^m$03$^{s}$.93 & $+$03$^\circ$31$^\prime$59$^{\prime\prime}$.99$^b$&0.017&67.4&0.33 \\
&&&&& \\ \hline
\end{tabular}
\label{source_info}
\end{table*}

\begin{table*}[ht]
\caption{Parameters of the VLBI observations.}
\begin{tabular}{lll}
\hline \\
Observing date & : & February 18, 1998 \\
Stations & : & 10-element~VLBA, Phased VLA, Effelsberg, Noto and Torun \\
Frequency & : & 4964.474~MHz \\
Bandwidth & : & 64~MHz \\
Polarization & : & LCP only \\
Scan length &: & 8~min 48~s \\
Aggregate bit rate & : & 128 Mbit~s$^{-1}$
 (8 baseband channels at 16 MSamples~s$^{-1}$ of 1 bit samples) \\
Fringe Finders & : & 3C~84, 3C~345, DA~193, and 4C~39.25 \\
Correlator & : & NRAO, Socorro, NM, USA \\
\\
\hline
\end{tabular}
\label{vlbi_log}
\end{table*}

Short observations of four compact bright sources, \viz,
3C~84, 3C~345, DA~193, and 4C~39.25
were included in the VLBI schedule as `fringe finders',
\ie strong sources intended to provide checks
of the basic parameters of the experiment, such as the set-up at each station,
station coordinates, source coordinates, etc.

Our Seyfert galaxy targets were too weak to be used to phase the
VLA array. Therefore, we phased the VLA with observations
of sources from the VLA calibrator list located near the targets
just prior to each VLBI scan. Since the VLA was not
typically the slowest slewing antenna in the array, it was usually possible
to phase the VLA while the slower antennas were still slewing,
minimizing the need to extend the phasing time into the start of the
VLBI scans. These phasing sources also served as VLA
phase calibrators during the calibration of VLA observations
in aperture-synthesis mode; see Sect.~\ref {aux_vla}.

We observed our targets in snap-shot mode, sequencing them in order
to (i) roughly equalize the number of scans for the various targets,
(ii) optimize the ($u,v$) coverage for each target and (iii) include
as many stations as possible in as many scans as possible for each target.
We used the SCHED software from NRAO \citep{Walker_SCHED} to create 
the schedule of observations.  
The duration of each scan was 8~min~48~s, and the number of scans 
per target ranged between 7 and 12.

\subsection{VLA observations}
\label {aux_vla}

During all of the VLBI scans in which the VLA participated,
simultaneous VLA data for the target in aperture-synthesis mode
were also acquired. This resulted in simultaneous arcsec-scale data
for all our targets. The calibrators used to phase
the VLA prior to the VLBI scans were chosen to be `P'
(unresolved) or `S' (slightly resolved) calibrators close to the targets.
As is noted above, these phase calibrators (listed in Table~\ref{vla_param})
served the dual purpose of phasing the VLA for the VLBI
observations and acting as phase calibrators for the aperture-synthesis-mode
VLA data. Typically, about 8.5~min of aperture-synthesis mode
VLA data were obtained for the target during each scan, as well
as $\sim$~2~min on the corresponding phase calibrator prior to each scan.
The centre wavelength and bandwidth of the observations are the
same as that of the VLBI observations.

\section{Data reduction}
\label{data_red}

\subsection{VLBI data}
\label{vlbi_red}

The data recorded at each station were correlated at the VLBA
correlator at NRAO, Socorro. The correlated data were reduced in the
standard way using the NRAO Astronomical Image Processing
System (AIPS) package. The amplitude calibration for all antennas
except for the VLA was carried out using gain curves supplied
by each station and system temperatures measured before or during
scan in each baseband channel. The amplitudes for baselines involving
the VLA were calibrated using flux densities for the targets
determined from the VLA data in aperture-synthesis mode
together with ratios of the VLA source-to-system temperature
measured during the observations. The resulting errors in the visibility
amplitudes are about 5~\% or less, as indicated by the data for the
compact fringe-finder DA~193 (whose flux density was 5.38~Jy).

The phases in the eight baseband channels were aligned by using the
AIPS task FRING to solve for the phases and delays for
a short observation of one of the fringe-finders and applying these solutions
to the data for all sources (so-called `manual phase calibration').
FRING was then used to solve for small residual phase and delays for all
the target scans, using the full bandwidth provided by the eight baseband
channels to improve the signal-to-noise ratio for the solutions.

Before imaging, the data were averaged over the entire frequency
bandwidth, then over time (with an averaging time of 30~s).
During the iterative imaging procedure, the antenna gains were
self-calibrated along with the source structure (hybrid mapping). We
were able to  iteratively self-calibrate only the phases
for all the targets; our attempts at amplitude self-calibration failed
due to the weakness of the targets and the wide range of baseline
sensitivities in the inhomogeneous array. The phased VLA
was used as the reference antenna in all cases, and uniform weighting
was used in order to maximize the angular resolution of the images.
The r.m.s. noise levels in the final images after completion of
the self-calibration process were 0.09~mJy~beam$^{-1}$
to 0.23~mJy~beam$^{-1}$ (Table~\ref{vlbi_param}).

We model-fit the final self-calibrated data using the Brandeis package
\citep{Gabuzda88}, 
which fits the data in the visibility domain.  This  
avoids  the need to account for the effect of convolution with the synthesized
beam in the image. The distribution of clean components was used to arrive 
at an initial ``seed'' model with up to five components (circular Gaussians or
point sources), with each component corresponding to one cluster of CLEAN 
components.  The components that 
emerged with significant emission as a result of the first iteration of the
model-fitting procedure were retained and fed into a second iteration in
order to arrive at the final result.  The results are listed in 
Table~\ref{vlbi_res}.  

We also list the total VLBI flux densities 
derived from the data in the image plane using the AIPS task IMEAN. 
(The task JMFIT, which fits two-dimensional Gaussians to components in the 
image plane gives very similar results to IMEAN in cases when there are no 
weak multiple components.) In the case of objects with significant extended 
emission, the flux density accounted for by model-fitting the visibilities 
turns out to be some what less than that obtained from the image, but
the models contained $> 80~\%$ of the total flux density in but a few
sources (see notes on individual sources below).

\subsection{VLA data}

The VLA data acquired in aperture-synthesis mode were processed
in the standard way using AIPS.  1331+305 (3C~286) was used as the
primary flux density calibrator. The VLA images were obtained using
uniform weighting. The sources were subjected to several cycles of
self-calibration, until the r.m.s. noise approached the thermal noise
levels. The typical 1~$\sigma$ noise levels in the
maps are in the range 30 ~$\mu$Jy~beam$^{-1}$
to 120~$\mu$Jy~beam$^{-1}$.
We used data only from the first IF channel (\ie only half
the total bandwidth), as the second IF channel suffered from unacceptably
high  noise levels.
For each image, the flux densities of the components that were
easily separable from
neighbouring components or extensions were derived by fitting
the data in the image plane with two-dimensional Gaussians
 using the AIPS task JMFIT.
Thus diffuse extensions/components were
left out of these estimates. The total flux density detected was
estimated by summing all the flux density in a specified
rectangular region using the AIPS task IMEAN.
The total errors in the flux densities are typically 5~\%, arising from the
calibration and the r.m.s. noise. 

\section{Results}
\label{res_an}

\subsection{The radio images}

The VLBI and VLA contour maps for all the targets
are shown in Fig.~\ref{maps1}. The sequence of maps is
ordered in right ascension. The spatial coordinates for the VLA
maps are given in absolute units, whereas coordinates for the VLBI
maps are given relative to the phase centre, since absolute position
information was lost during the self-calibration. An ellipse in a box
in the lower left-hand corner of each map shows the shape of the synthesized
beam FWHM. A cross in each of the VLA maps marks the position of
the optical nucleus, with the size of the cross representing the uncertainty
in the position. The optical positions are from
\cite{Gallouet75}, \cite{Clements81,Clements83}, and
\cite{ArgyleEldridge90}.  \cite{Gallouet75} quote typical position
errors of 6~arcsec, and errors of 0.09~arcsec to 0.2~arcsec
are given by the other authors.  All positions are given in J2000 coordinates.

The beam FWHM, beam position angle, peak surface brightness,
r.m.s.~noise level and contour levels for all the maps are shown in
Table~\ref{vla_param} (VLA maps) and Table~\ref{vlbi_param}
(VLBI maps).
The r.m.s. noise level (1~$\sigma$) in the images
was estimated using regions that were free of source emission using the
AIPS task IMEAN. The r.m.s. noise of an empty region was
then fed into a second iteration of IMEAN for the entire image, to
obtain the r.m.s. deviation from a fit to the Gaussian (noise) part of
the pixel histogram for the whole image.

\begin{table*}[]
\caption{VLA phase calibrators and the parameters pertaining to the VLA maps.}
\begin{tabular}{lcccccl} \hline \\
Object    &Phase&  \multicolumn{2}{c}{CLEAN beam} &   Peak\hspace{0.8mm}   &   Image &      Contours  \\ \cline{3-4}
          &calibrator&  & P.A. & &r.m.s. & \\
          && (arcsec$^2$)&(deg)&\multicolumn{2}{c}{(mJy~beam$^{-1}$) ($\mu$Jy~beam$^{-1}$)}&(\% of peak surface brightness) \\
\\ \hline
\\
  Mrk 1     &  0119$+$321           & 0.65 $\times$ 0.49& 54.0 & 24.08 &35& $-$0.3, 0.8, 1, 2, 4, 6, 8, 10, 12, 16, 24, 32, 40, 48, 56, 64, 80, 90 \\
  MCG 8$-$11$-$11& 0604$+$442   & 0.84 $\times$ 0.44& 57.7 & 23.57 &52&$-$0.6, 0.6, 1, 2, 4, 6, 8, 10, 12, 16, 24, 32, 40, 48, 56, 64, 80, 90 \\
  NGC 2273  &  0650$+$600         & 0.70 $\times$ 0.46& 60.2 & 12.13 &85   &$-$1.6, 1.6, 2, 4, 6, 8, 10, 12, 16, 24, 32, 40, 48, 56, 64, 80, 90 \\
  Mrk 78    &  0737$+$596       & 0.70 $\times$ 0.52& 71.6 &  8.87 &30&  $-$1, 1, 1.6, 2, 4, 6, 8, 10, 12, 16, 24, 32, 40, 48, 56, 64, 80, 90 \\
  Mrk 1218  &  0837$+$249       & 0.53 $\times$ 0.41&$-$68.3 & 18.72 &121& $-$1.2, 1.2, 2, 4, 6, 8, 10, 12, 16, 24, 32, 40, 48, 56, 64, 80, 90 \\
  NGC 2639  &  0832$+$492       & 0.64 $\times$ 0.45& 81.6 & 63.47 &70& $-$0.4, 0.4, 1, 2, 4, 6, 8, 10, 12, 16, 24, 32, 40, 48, 56, 64, 80, 90 \\
  Mrk 766   &  1217$+$301       & 0.65 $\times$ 0.48&$-$71.9 & 13.75 &35& $-$0.8, 1, 2, 4, 6, 8, 10, 12, 16, 24, 32, 40, 48, 56, 64, 80, 90 \\
  Mrk 477   &  1419$+$543       & 0.99 $\times$ 0.55&$-$63.3 & 22.97 &34& $-$0.3, 0.8, 1, 2, 4, 6, 8, 10, 12, 16, 24, 32, 40, 48, 56, 64, 80, 90 \\
  NGC 5929  &  1458$+$373       & 1.16 $\times$ 0.98& 75.4 & 21.51 &95& $-$0.8, 1.6, 2, 4, 6, 8, 10, 12, 16, 24, 32, 40, 48, 56, 64, 80, 90 \\
  NGC 7212  &  2151$+$071       & 1.18 $\times$ 0.68&$-$65.1 & 19.59 &77& $-$0.8, 0.8, 1, 2, 4, 6, 8, 10, 12, 16, 24, 32, 40, 48, 56, 64, 80, 90 \\
  Ark 564   &  2236$+$284       & 0.85 $\times$ 0.75&  8.4 &  8.62 &40& $-$1.6, 1.6, 2, 4, 6, 8, 10, 12, 16, 24, 32, 40, 48, 56, 64, 80, 90 \\
  NGC 7469  &  2257$+$077       & 0.61 $\times$ 0.55& 54.9 & 21.97 &49& $-$0.6, 0.6, 1, 2, 4, 6, 8, 10, 12, 16, 24, 32, 40, 48, 56, 64, 80, 90 \\
  Mrk 530   &  2323$-$032       & 0.82 $\times$ 0.53& 35.2 &  8.02 &41& $-$1.6, 1.6, 2, 4, 6, 8, 10, 12, 16, 24, 32, 40, 48, 56, 64, 80, 90 \\
  Mrk 533   &  2330$+$110       & 0.82 $\times$ 0.53& 35.2 & 38.19 &72& $-$0.4, 0.4, 1, 2, 4, 8, 15, 24, 32, 40, 48, 56, 64, 72, 80, 90 \\
  NGC 7682  &  2330$+$110       & 0.94 $\times$ 0.64& 38.2 & 21.98 &36& $-$0.4, 0.6, 1, 2, 4, 6, 8, 10, 12, 16, 24, 32, 40, 48, 56, 64, 80, 90 \\
\\
\hline \\
\end{tabular}
\label{vla_param}
\end{table*}

\begin{table*}[]
\caption{Contour levels and other parameters for VLBI maps.
$\star$:
Some scans did not include all the stations of the global array.}
\begin{tabular}{lcccccl} \hline \\
Object& No. of&  \multicolumn{2}{c}{CLEAN beam} &   Peak   &   Image &      Contours  \\\cline{3-4}
    & scans$^\star$&  & P.A. & &r.m.s. & \\
 && (mas$^2$)&(deg)&\multicolumn{2}{c}{(mJy~beam$^{-1}$) ($\mu$Jy~beam$^{-1}$)}&(\% of peak surface brightness) \\ \\
\hline \\
Mrk 1       &13& 3.69 $\times$ 1.78&$-$12.2 & 6.7 & 135& $-$6, 8, 10, 12, 16, 24, 32, 40, 48, 56, 64, 80, 90 \\
MCG 8$-$11$-$11 &13& 2.08 $\times$ 1.60&$-$53.9 & 6.9 & 121& $-$8, 8, 10, 12, 16, 24, 32, 40, 48, 56, 64, 80, 90 \\
NGC 2273    &11& 3.02 $\times$ 2.52& 28.1 & 7.5 & 131& $-$6, 8, 10, 12, 16, 24, 32, 40, 48, 56, 64, 80, 90 \\
Mrk 78      &18& 3.61 $\times$ 2.43& 85.0 & 7.1 & 145& $-$8, 8, 10, 12, 16, 24, 32, 40, 48, 56, 64, 80, 90 \\
Mrk 1218    & 7& 2.88 $\times$ 1.05&$-$10.4 & 6.7 & 110& $-$5, 6, 8, 10, 12, 16, 24, 32, 40, 48, 56, 64, 80, 90 \\
NGC 2639    &12& 1.82 $\times$ 1.08& $-$6.8 &34.8 & ~85 & $-$0.5, 0.5, 1, 2, 4, 8, 16, 24, 32, 40, 48, 56, 64, 72, 80, 90 \\
Mrk 766     &11& 2.78 $\times$ 1.65& 18.3 & 7.1 & 131& $-$8, 8, 10, 12, 16, 24, 32, 40, 48, 56, 64, 80, 90 \\
Mrk 477     &10& 4.48 $\times$ 3.26& 50.0 & 9.4 & 144& $-$4, 4, 6, 8, 10, 12, 16, 24, 32, 40, 48, 56, 64, 80, 90 \\
NGC 5929    &11& 2.91 $\times$ 1.80& 31.5 & 6.9 & 120& $-$6, 8, 10, 12, 16, 24, 32, 40, 48, 56, 64, 80, 90 \\
NGC 7212    & 7& 4.27 $\times$ 2.65& $-$3.5 & 7.2 & 233& $-$8, 8, 10, 12, 16, 24, 32, 40, 48, 56, 64, 80, 90 \\
Ark 564     &11& 2.54 $\times$ 1.25& 13.0 & 3.7 & ~91& $-$16, 12, 16, 24, 32, 40, 48, 56, 64, 80, 90 \\
NGC 7469    & 7& 3.65 $\times$ 2.67&  2.7 & 7.4 & 146& $-$8, 8, 10, 12, 16, 24, 32, 40, 48, 56, 64, 80, 90 \\
Mrk 530     & 7& 3.93 $\times$ 2.54& $-$4.8 & 8.8 & 132& $-$6, 6, 8, 10, 12, 16, 24, 32, 40, 48, 56, 64, 80, 90 \\
Mrk 533     & 7& 6.22 $\times$ 3.89& $-$4.2 &10.2 & 117& $-$4, 4, 6, 8, 10, 12, 16, 24, 32, 40, 48, 56, 64, 80, 90 \\
NGC 7682    & 7& 4.00 $\times$ 3.00&  1.9 & 8.9 & 149& $-$6, 8, 10, 12, 16, 24, 32, 40, 48, 56, 64, 80, 90 \\ \\ \hline \\
\end{tabular}
\label{vlbi_param}
\end{table*}

Table~\ref{vlbi_res} summarizes the VLBI observational results.
It gives the object name, {\it a priori} coordinates used for the
VLBI observations and for correlation of the interferometric data, 
peak surface brightness,
components sizes, the total (summed) flux density for the model components
and the total flux density in the image obtained using the AIPS task IMEAN.

\begin{table*}[]
\caption{The VLBI results.
$\dag$: See description of the Seyfert in Sect.~\ref{descrip_sour}.}
\begin{tabular}{l|cr|c|ccccc|c} \hline
 & \multicolumn{2}{|c|}{{\it a priori} position used}& &\multicolumn{5}{|c|}{Model-fit components}& Total \\ \cline{5-9}
 &\multicolumn{2}{|c|}{for the correlation}        &Peak surface &\multicolumn{2}{c}{Position relative}    &FWHM & Flux & Total & flux density \\ \cline{2-3}
  Object    & \multicolumn{2}{r|} {Right Ascention ~~~Declination} &brightness &  \multicolumn{2}{c}{to phase centre} && density  & & from the image  \\
 & \multicolumn{2}{c|}{(J2000)}  &(mJy beam$^{-1}$) &$\alpha$(mas)&$\delta$(mas)&(mas)&(mJy)&(mJy)&(mJy) \\ \hline
&&&&&&&&& \\
 Mrk 1      &01$^h$16$^m$07$^{s}$.22 & $+$33$^\circ$05$^\prime$21$^{\prime\prime}$.6 &6.7 & $+$0.0 & $+$0.0 & 1.0  & 4.4 & 4.4 & 6.9 \\
 MGC 8-11-11&05$^h$54$^m$53$^{s}$.61 & $+$46$^\circ$26$^\prime$21$^{\prime\prime}$.7 &6.9 & $+$0.0 & $+$0.0 & 0.8  & 4.9 & 4.9 & 7.8 \\
 NGC 2273   &06$^h$50$^m$08$^{s}$.64 & $+$60$^\circ$50$^\prime$44$^{\prime\prime}$.9 &7.5 & $+$0.0 & $-$0.1 & 1.2  & 6.3 & 6.3 & 7.7 \\
 Mrk 78     &07$^h$42$^m$41$^{s}$.74 & $+$65$^\circ$10$^\prime$37$^{\prime\prime}$.8 &7.1 & $+$0.0 & $+$0.1 & 0.7  & 5.2 & 8.5 & 9.0 \\
            &                                &                                             & & $-$0.7 & $+$0.1 & 0.2  & 1.4 &     & \\
            &                                &                                             & & $-$1.5 & $-$2.0 & 0.6  & 1.0 &     & \\
            &                                &                                             & & $+$0.8 & $-$3.4 & 0.2  & 0.3 &     & \\
            &                                &                                             & & $+$3.7 & $+$5.3 & 0.6  & 0.6 &     & \\
 Mrk 1218   &08$^h$38$^m$10$^{s}$.95 & $+$24$^\circ$53$^\prime$42$^{\prime\prime}$.9 &6.7 & $+$0.0 & $+$0.0 & 1.1  & 7.0 &12.3$^\dag$ &14.8 \\
            &                                &                                             & & $+$0.1 & $-$5.9 & 2.8  & 1.7 &     & \\
            &                                &                                             & & $-$1.8 & $+$4.3 & 2.1  & 2.8 &     & \\
            &                                &                                             & & $+$6.5 & $-$5.3 & 0.9  & 0.8 &     & \\
 NGC 2639   &08$^h$43$^m$38$^{s}$.07 & $+$50$^\circ$12$^\prime$20$^{\prime\prime}$.1 &34.8 & $+$0.0 & $+$0.0 & 0.1  &31.6 &39.5 &40.2 \\
            &                                &                                             & & $+$0.5 & $-$0.1 & 0.3  & 8.0 &     & \\
 Mrk 766    &12$^h$18$^m$26$^{s}$.52 & $+$29$^\circ$48$^\prime$46$^{\prime\prime}$.5 &7.1 & $+$0.0 & $+$0.0 & 0.9  & 3.7 & 3.6 & 6.3 \\
 Mrk 477    &14$^h$40$^m$38$^{s}$.10 & $+$53$^\circ$30$^\prime$16$^{\prime\prime}$.1 &9.4 & $+$0.3 & $+$0.0 & 1.4  & 8.2 & 8.1 &10.3 \\
 NGC 5929   &15$^h$26$^m$06$^{s}$.17 & $+$41$^\circ$40$^\prime$14$^{\prime\prime}$.4 &6.9 & $+$0.0 & $-$0.2 & 0.8  & 6.0 & 6.2 & 7.2 \\
            &                                &                                             & & $+$3.8 & $+$2.1 & 0.2  & 0.3 &     & \\
 NGC 7212   &22$^h$07$^m$02$^{s}$.06 & $+$10$^\circ$14$^\prime$02$^{\prime\prime}$.6 &7.2 & $+$0.0 & $-$0.1 & 1.1  & 6.5 & 7.0 & 7.9 \\
            &                                &                                             & & $-$5.0 & $-$1.0 & 1.1  & 0.5 &     & \\
 Ark 564    &22$^h$42$^m$39$^{s}$.36 & $+$29$^\circ$43$^\prime$30$^{\prime\prime}$.9 &3.7 & $+$0.2 & $-$0.1 & 0.9  & 3.2 & 3.2$^\dag$ & 9.1 \\
 NGC 7469   &23$^h$03$^m$15$^{s}$.62 & $+$08$^\circ$52$^\prime$26$^{\prime\prime}$.1 &7.4 & $+$0.1 & $-$0.1 & 0.9  & 5.5 & 6.1 & 7.4 \\
            &                                &                                             & & $+$0.0 & $-$4.5 & 0.1  & 0.7 &     & \\
 Mrk 530    &23$^h$18$^m$56$^{s}$.65 & $+$00$^\circ$14$^\prime$38$^{\prime\prime}$.0 &8.8 & $+$0.0 & $+$0.5 & 0.9  & 7.6 & 8.6 & 9.7 \\
            &                                &                                             & & $-$1.3 & $-$1.4 & 0.1  & 0.3 &     & \\
            &                                &                                             & & $+$3.4 & $-$1.1 & 0.9  & 0.8 &     & \\
 Mrk 533    &23$^h$27$^m$56$^{s}$.71 & $+$08$^\circ$46$^\prime$44$^{\prime\prime}$.1 &10.2 & $+$0.1 & $-$0.1 & 1.1  & 9.7 &16.9 &20.5 \\
            &                                &                                             & & $+$6.2 & $-$2.0 & 0.7  & 2.2 &     & \\
            &                                &                                             & &$+$16.7 & $-$4.0 & 0.5  & 1.2 &     & \\
            &                                &                                             & & $-$6.0 & $+$0.4 & 2.2  & 2.3 &     & \\
            &                                &                                             & &$-$13.8 & $-$0.3 & 0.8  & 1.6 &     & \\
 NGC 7682   &23$^h$29$^m$03$^{s}$.92 & $+$03$^\circ$31$^\prime$59$^{\prime\prime}$.9 &8.9 & $+$0.1 & $-$0.3 & 0.9  & 7.6 &11.5$^\dag$ &14.2 \\
            &                                &                                             & & $+$3.2 & $-$2.5 & Point& 2.1 &     & \\
            &                                &                                             & & $-$0.1 & $-$5.5 & Point& 1.8 &     & \\
&&&&&&&&& \\
 \hline
\end{tabular}
\label{vlbi_res}
\end{table*}

The results of the VLA imaging are summarized in Table~\ref{vla_res},
which gives the object name, radio position of each
component, offset of radio position from optical position,
peak surface brightness,
total flux density of each component, and the largest angular
size of the source.  The component flux densities were 
derived by model-fitting Gaussians in the image plane to 
those components that were easily separable from
neighbouring components or extensions 
using the AIPS task JMFIT. For multiple components the corresponding
parameters are given in separate rows of  Table~\ref{vla_res}.
The total flux density is that in a rectangular region
enclosing all of the detected emission, derived 
using the AIPS task IMEAN in every case. The model-fitting procedure does not 
account for all the extended emission since only Gaussian components were 
used, and therefore the sum of the component flux densities is often lower than 
the total flux density. The largest angular size for each source is the
size of the contour corresponding to 
5~\% of the peak surface brightness in the case of extended sources,
which, in all cases is well above the noise in the image. For the
unresolved sources, the FWHM of the major axis of the synthesized beam is 
listed as the upper limit.

\begin{table*}[]
\caption{VLA observational results.
$\ddag$: The peak surface brightness of the radio image of
NGC~5929 coincides with one of the hotspots and not the nucleus.}
\begin{tabular}{l|cc|cc|c|cc|c}
\hline
           & \multicolumn{2}{|c|}{Position of radio peak} &\multicolumn{2}{|c|}{optical-radio offsets} & Peak surface  & Component & Total &  Largest \\ \cline{2-5}
 Object &  \multicolumn{2}{|r|} {Right Ascension ~~~~~~~Declination} & \multicolumn{2}{|r|} {$\Delta$ RA ~~~~~~~~~$\Delta$ Dec}& brightness &\multicolumn{2}{c|}{flux density} & angular size \\
 &\multicolumn{2}{c|}{(J2000)}& \multicolumn{2}{|c|} {(arcsec)} &(mJy beam$^{-1}$) &\multicolumn{2}{c|}{(mJy)}&(arcsec) \\ \hline
&&&&&&&& \\
Mrk 1      &01$^h$16$^m$07$^{s}$.209&$+$33$^\circ$05$^\prime$21$^{\prime\prime}$.66&$+$0.61 &$-$0.71 &24.08&27.1& 27.8& $<$~0.65 \\
MCG 8-11-11&05$^h$54$^m$53$^{s}$.620&$+$46$^\circ$26$^\prime$21$^{\prime\prime}$.41&$-$0.15 &$+$0.20 &23.57&65.3 &78.0& 3.48 \\
NGC 2273   &06$^h$50$^m$08$^{s}$.646&$+$60$^\circ$50$^\prime$44$^{\prime\prime}$.94&$+$1.11 &$+$0.07 &12.13&14.3 &22.0& 2.25 \\
 &06$^h$50$^m$08$^{s}$.534&$+$60$^\circ$50$^\prime$45$^{\prime\prime}$.08& & &&6.3 &&  \\
Mrk 78     &07$^h$42$^m$41$^{s}$.724&$+$65$^\circ$10$^\prime$37$^{\prime\prime}$.44&$+$0.09 &$+$0.02 & 8.87&10.1  &13.1& 3.88 \\
           &07$^h$42$^m$41$^{s}$.340&$+$65$^\circ$10$^\prime$37$^{\prime\prime}$.63& & &&2.3  &&   \\
Mrk 1218   &08$^h$38$^m$10$^{s}$.945&$+$24$^\circ$53$^\prime$42$^{\prime\prime}$.82&$+$0.08 &$+$0.10 &18.72&19.7 &27.7& 1.28 \\
NGC 2639   &08$^h$43$^m$38$^{s}$.072&$+$50$^\circ$12$^\prime$19$^{\prime\prime}$.97&$-$0.08 &$+$0.35 &63.47&80.5 &85.8& $<$~0.65 \\
Mrk 766    &12$^h$18$^m$26$^{s}$.517&$+$29$^\circ$48$^\prime$46$^{\prime\prime}$.50&$-$0.11 &$+$0.16 &13.75&16.8 &17.9& $<$~0.65 \\
Mrk 477    &14$^h$40$^m$38$^{s}$.097&$+$53$^\circ$30$^\prime$15$^{\prime\prime}$.96&$+$0.20 &$+$0.04 &22.97&26.6 &27.3& $<$~0.99 \\
NGC 5929   &15$^h$26$^m$06$^{s}$.113&$+$41$^\circ$40$^\prime$14$^{\prime\prime}$.02$^\ddag$&$+$0.30 &$+$0.70 &21.51&25.5 &34.4& 3.47 \\
NGC 7212   &22$^h$07$^m$01$^{s}$.998&$+$10$^\circ$14$^\prime$00$^{\prime\prime}$.69&$+$0.18 &$-$0.35 &19.59&30.0 &31.0& 3.09 \\
Ark 564    &22$^h$42$^m$39$^{s}$.332&$+$29$^\circ$43$^\prime$31$^{\prime\prime}$.07&$+$0.01 &$+$0.24 &8.62&10.7 &11.4& $<$~0.85 \\
NGC 7469   &23$^h$03$^m$15$^{s}$.616&$+$08$^\circ$52$^\prime$26$^{\prime\prime}$.02&$+$0.06 &$+$0.37 &21.97&26.9&47.9& 2.76 \\
Mrk 530    &23$^h$18$^m$56$^{s}$.653&$+$00$^\circ$14$^\prime$37$^{\prime\prime}$.96&$-$0.50 &$+$0.27 &8.02&8.3 & 10.2& $<$~0.82 \\
Mrk 533    &23$^h$27$^m$56$^{s}$.712&$+$08$^\circ$46$^\prime$44$^{\prime\prime}$.13&$+$0.12 &$+$0.40 &38.19&60.4 &58.8& 1.89 \\
NGC 7682   &23$^h$29$^m$03$^{s}$.918&$+$03$^\circ$31$^\prime$59$^{\prime\prime}$.92&$+$0.18 &$+$0.07 &21.98&22.3 &22.6& $<$~0.94 \\
&&&&&&&& \\ \hline
\end{tabular}
\label{vla_res}
\end{table*}

\subsection{Description of the sources}
\label{descrip_sour}

We describe here the arcsec-scale radio structure measured by the VLA
(subsection ``{\em $Y$}") and the mas-scale structure derived from the global
VLBI observations (subsection ``{\em $G$}") for each of the Seyfert galaxies.
We also present two-frequency spectral indices, $\alpha$, when measurements
for other frequencies with similar resolution ($\simeq$ 1~arcsec) were
available in the literature (the spectral index is defined as
$S_{\nu}$~$\propto$~$\nu^{-\alpha}$).

Variability of the sources and differing
angular resolutions for the images used can cause large uncertainties in
the calculated spectral indices; we have accordingly attempted to use data
that were obtained as near in time as possible to ours, and with angular
resolution as close as possible to ours when calculating the spectral indices.

\medskip
\noindent {\bf Mrk 1} (0113+328)

\begin{enumerate}
\item[{\em {\small $Y$}:}]

Mrk 1 is unresolved. The total flux density
is similar to that presented by \cite{Ulvestad81}
and is less than the value of 38~mJy found by
\cite{SramekTovmassian75}. \cite{Ulvestad81}
suggest that, along with an unresolved component, there is a
faint ($\sim$~2~mJy) extension $\sim$~0.4~arcsec to the south.
This faint extension is not detected in our map. Since
\cite{Ulvestad81} do not show a map for this source, it is
difficult to judge whether we should have detected it.
The peak radio surface brightness in our map is displaced $\sim$~1~arcsec
to the south-west of the optical nucleus \citep{Clements81}.

\item[{\em {\small $G$}:}]

The image shows only a single somewhat resolved component.
\cite{Kukula99} detected a compact radio central component
surrounded by a halo of emission approximately 100~mas across
at 1.7~GHz using the EVN (angular resolution $\sim$ 20~mas).
They find evidence for weak emission extending to the south,
possibly leading into the larger structure
\citep{Ulvestad81}. However, they
do not find any evidence for linear structure
on scales larger than a few parsec. Our best model contains
about 64~\% of the total flux density on VLBI scales. The maximum extent of
the source in our image is $\sim$ 30~pc across.

\end{enumerate}

\noindent {\bf MCG 8$-$11$-$11} (0551+464)

\begin{enumerate}
\item[{\em {\small $Y$}:}]

Our map shows a strong nuclear component
with a compact elongation extending toward the
north-west (P.A.~$-$45$^{\circ}$). There is also larger-scale extended
emission towards the north. The bright central component is coincident
with the optical nucleus \citep{Clements81}, and the total extent of the
north-south structure is $\sim$~3~arcsec.  The radio image of
\cite{Schmitt01} at 8.4~GHz (VLA~$A$ array) is
similar to ours. The 15~GHz map of \cite{UlvestadWilson86}
resolves the central component into a triple structure.
The largest angular size of the source is $\sim$~3.5~arcsec.

\item[{\em {\small $G$}:}]

The image shows a single somewhat resolved component.
The weak component to the south-west (if real)
has a flux density of $\sim$~0.6~mJy. Our best model contains about
63~\% of the total flux density on VLBI scales.

\end{enumerate}

\noindent {\bf NGC 2273} (0645+609)

\begin{enumerate}
\item[{\em {\small $Y$}:}]

Our map has similar features to
those seen previously by \cite{Nagar99II} at 8.4~GHz and
\cite{UlvestadWilson84I} at 5~GHz. The
source consists of an unequal double with a separation of nearly
1~arcsec in P.A. $\sim$~$-$80$^{\circ}$. The WSRT image of this
source at 5~GHz reported by \cite{Baum93} shows  amorphous structure
on larger scales ($>$~2.5~arcsec). Our central-component flux density
and total flux density are slightly higher than those in the 5~GHz
image of \cite{UlvestadWilson84II} with a similar resolution. The optical
nucleus \citep{ArgyleEldridge90} is $\sim$~0.6~arcsec to the east of the
peak radio surface brightness.

\item[{\em {\small $G$}:}]

The image shows a single somewhat resolved component.
There is some evidence from the distribution of CLEAN
components for possible weak emission out to $\sim$~1~mas from the
phase centre.

\end{enumerate}

\noindent {\bf Mrk 78} (0737+652)

\begin{enumerate}
\item[{\em {\small $Y$}:}]

Mrk 78 is an extended source nearly 4~arcsec in size in
our 5~GHz image. The bright central component is
resolved and is coincident with the optical nucleus \citep{Clements81}.
The secondary component is nearly 2~arcsec to the west of the main peak
(P.A. $\sim$~$-$90$^{\circ}$). The brighter component is also extended
roughly in the east-west direction (P.A. $\sim$~80$^{\circ}$).

\item[{\em {\small $G$}:}]

Our map shows a faint extension towards the south-west
(P.A. $\sim$ $-$130$^{\circ}$).
Model-fitting the visibility data with five circular Gaussians 
accounts for almost all the flux density measured from the image.

\end{enumerate}

\noindent {\bf Mrk 1218} (0835+250)

\begin{enumerate}
\item[{\em {\small $Y$}:}]

\cite{Ulvestad86} found this object to be only slightly resolved
at 5~GHz (VLA~$A$-array) and 1.4~GHz
(VLA~$B$-array). Our map shows a diffuse,
faint ($\sim$~2~$\sigma$) extension toward the north-west in
P.A. $\sim$~$-$45$^{\circ}$. The total flux density of the source is
dominated by the bright central component, which is coincident with the optical
nucleus \citep{Clements83} within the errors. The total flux density
of 25.5~mJy measured by us is consistent with the value of 23.0~mJy
presented by \citep{Ulvestad86}. The spectral index derived from our
5~GHz observations and the 1.4~GHz observations of
\cite{Ulvestad86} (obtained in February 1984) is
$\alpha_{1.4~{\rm GHz}}^{5~{\rm GHz}} = 0.8$.

\item[{\em {\small $G$}:}]

 Apart from the brightest component (\abt~7~mJy), we detect several 
blobs of fainter emission, some in the approximate direction of the kpc-scale 
elongation. Model-fitting the visibilities with four circular Gaussians 
accounts for most of the emission measured in the image plane by
summing pixel values using the AIPS task IMEAN, and gives a total
flux density approximately the same as that obtained by fitting
Gaussians in the image plane using AIPS task JMFIT, \viz 12~mJy.

\end{enumerate}

\begin{figure*}[ht]
\centering
\includegraphics[height=6.4cm]{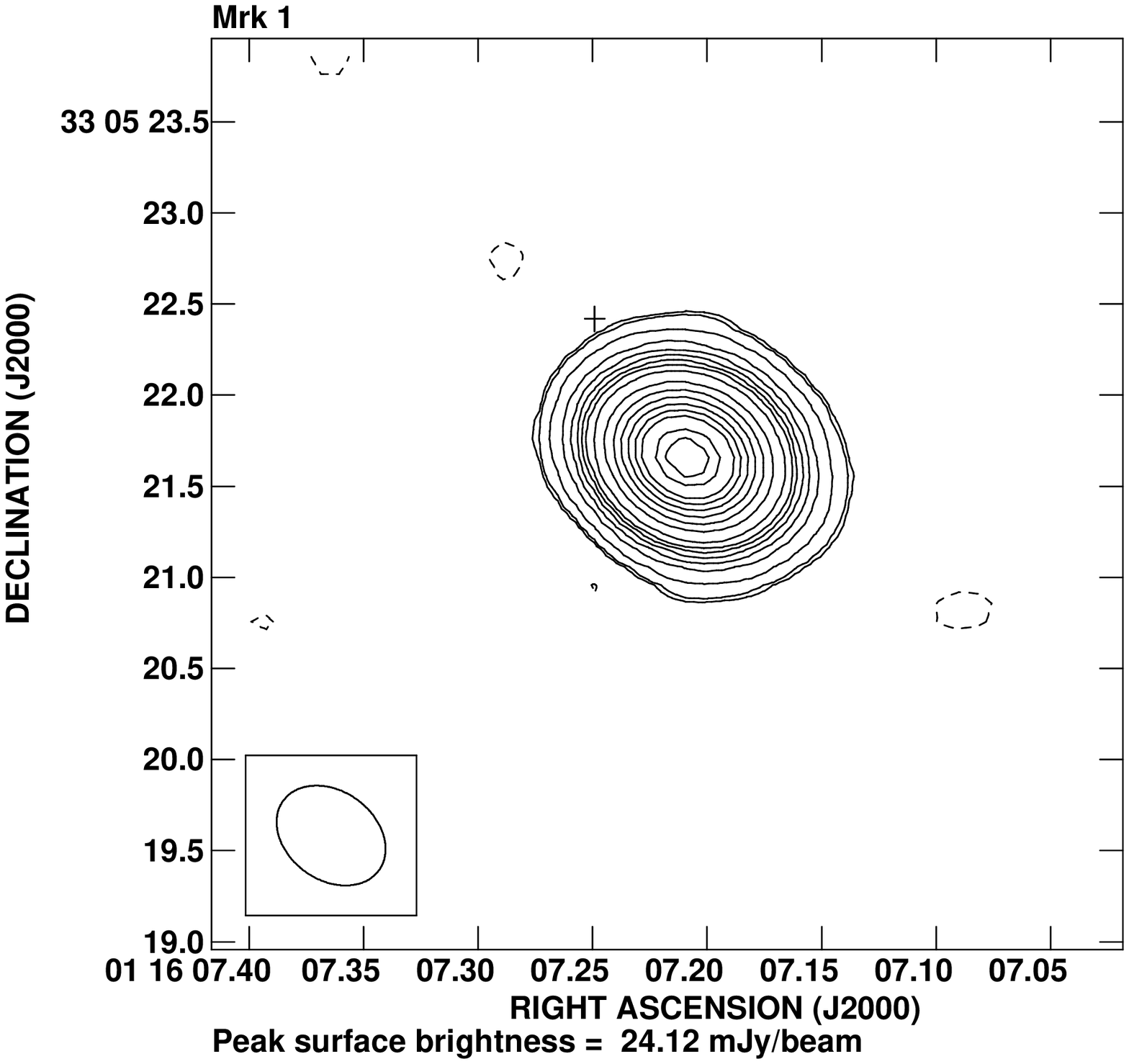} \hspace{1cm}
\includegraphics[height=6.4cm]{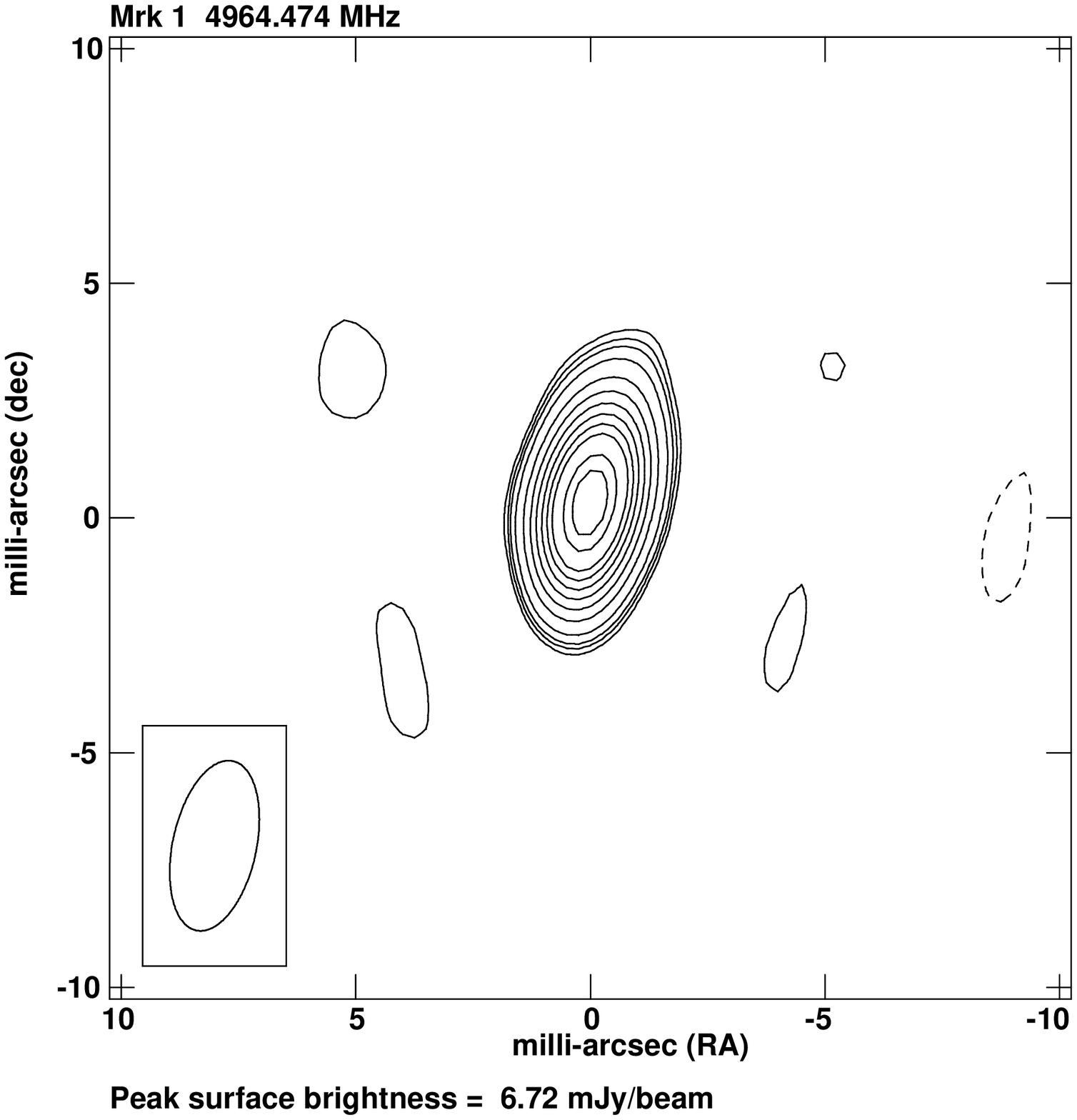} \\
\includegraphics[height=6.4cm]{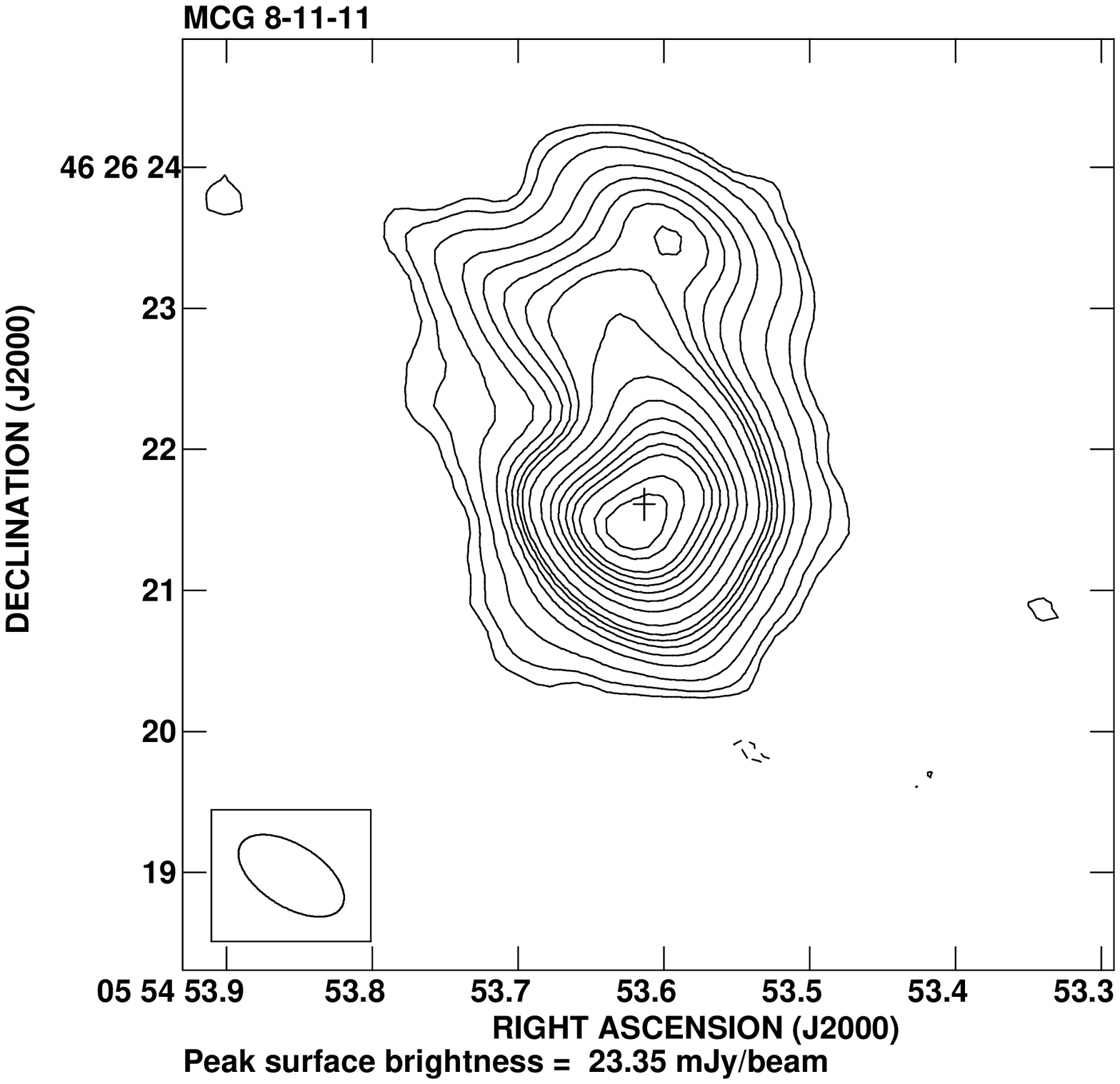} \hspace{1cm}
\includegraphics[height=6.4cm]{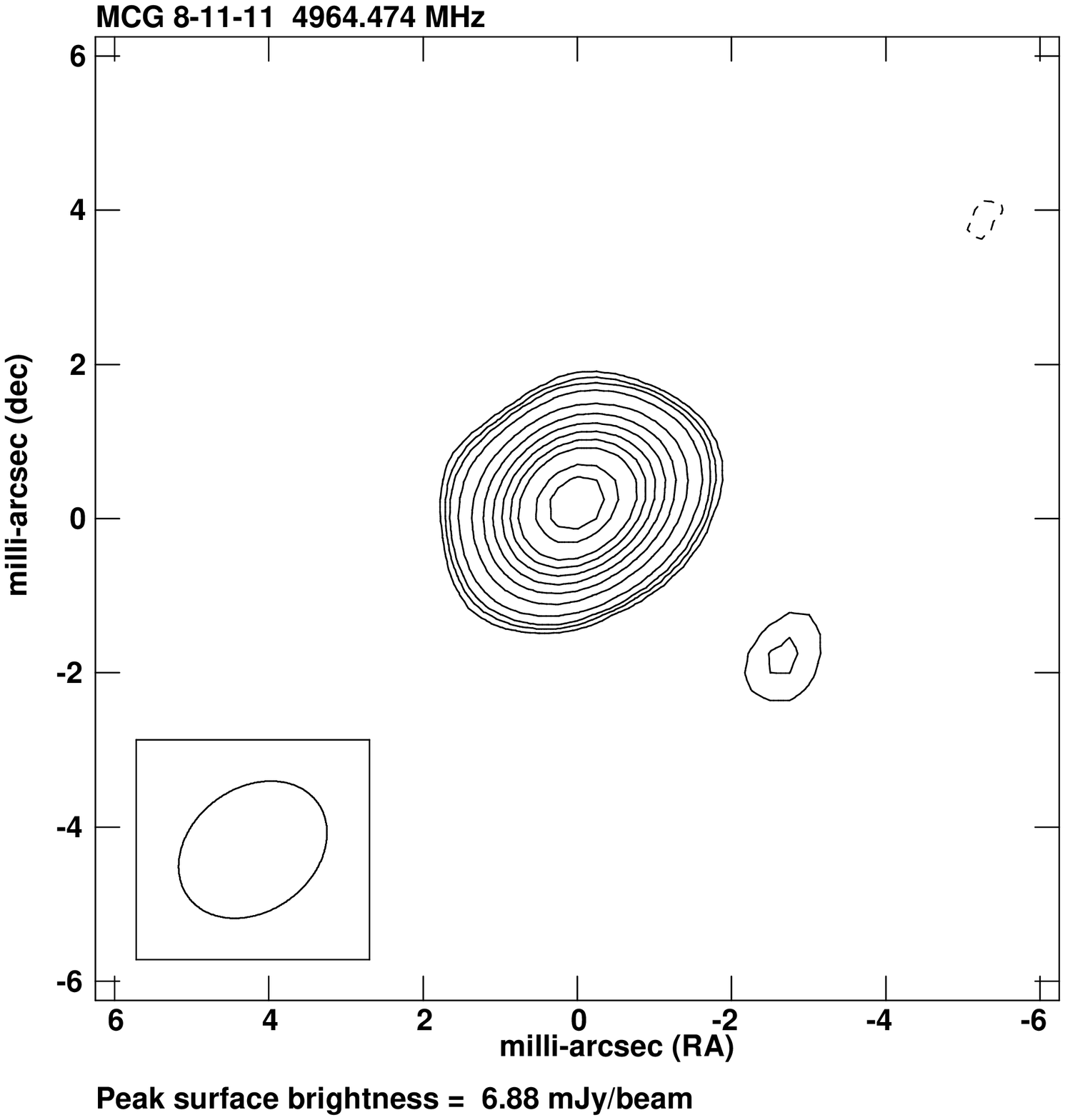} \\
\includegraphics[height=6.4cm]{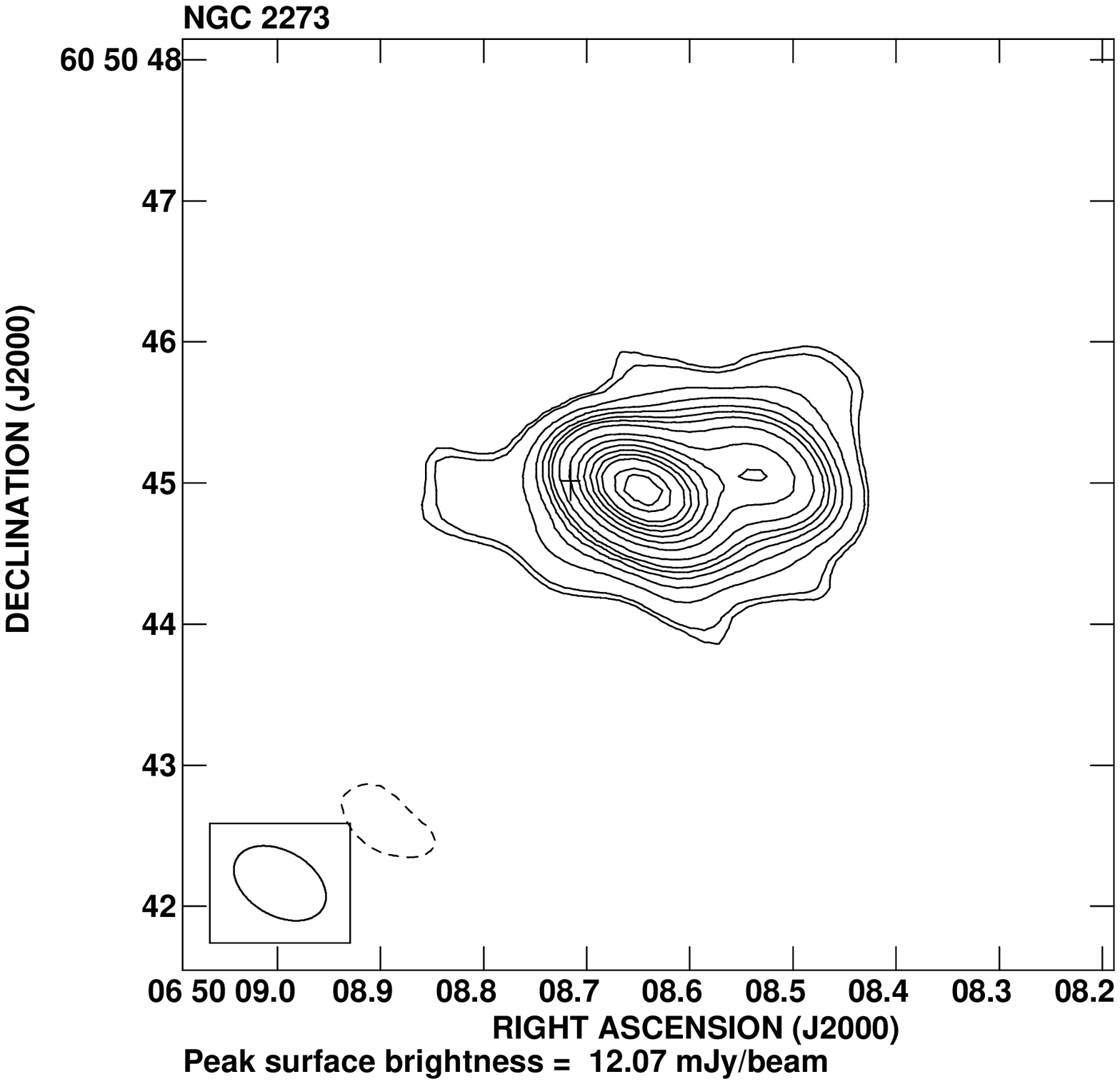} \hspace{1cm}
\includegraphics[height=6.4cm]{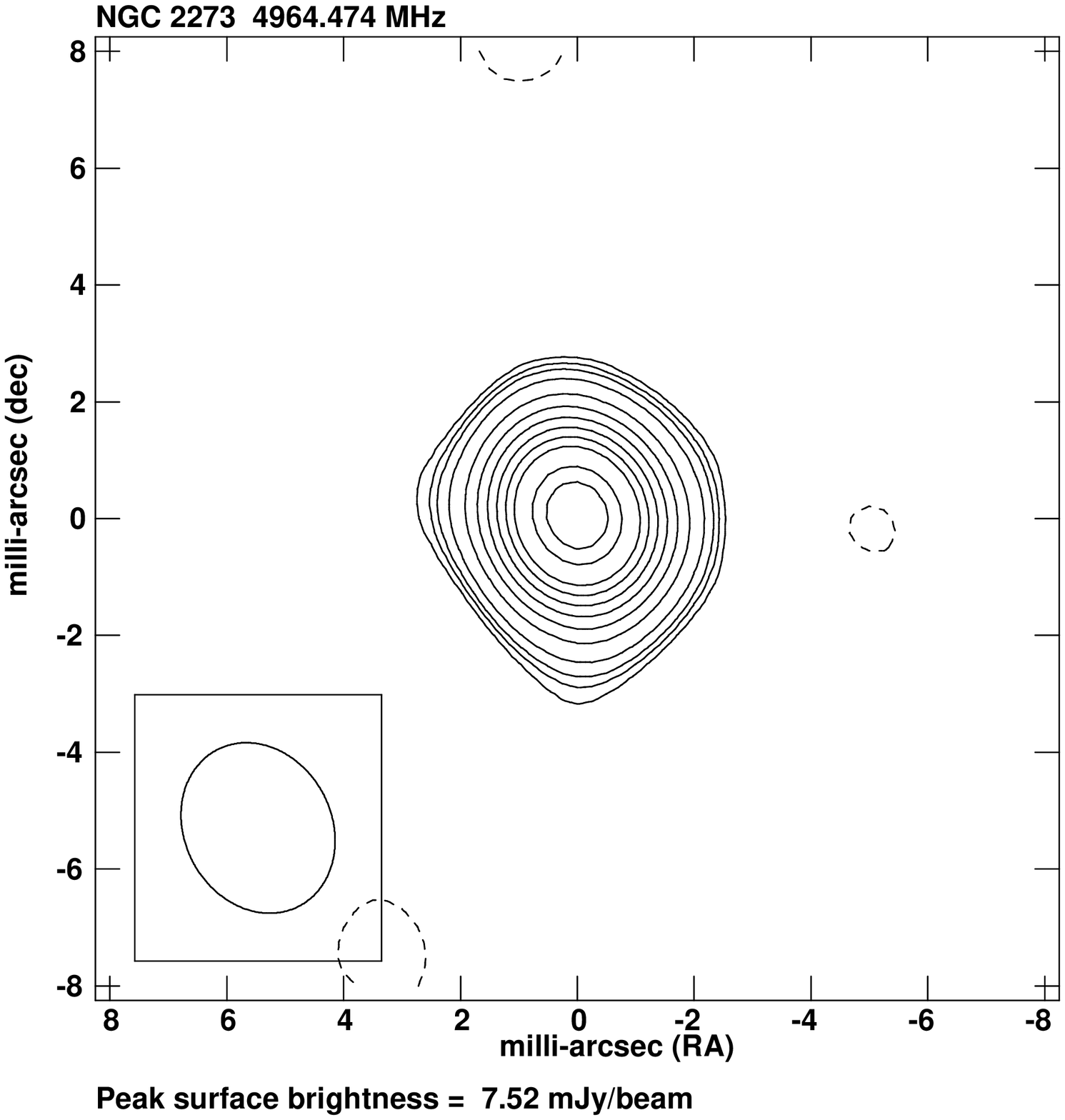} \\
\caption{The 5~GHz contour maps for all the targets, ordered in
right ascension. 
The VLA and VLBI maps are in the left
and right panels, respectively. 
The source name is given in the top left corner,
and the boxed ellipse in the lower left-hand corner shows the shape of the
FWHM synthesized beam in each case. A cross in the VLA maps
marks the position of the optical nucleus, with the size of the cross
representing the uncertainty in the position.  All positions are given in
J2000 coordinates. The contour levels and surface brightness peaks are
listed in Tables~\ref {vla_param} (VLA maps) and \ref {vlbi_param}
(VLBI maps)}.
\label{maps1}
\end{figure*}

\addtocounter{figure}{-1}
\begin{figure*}
\centering
\includegraphics[height=6.4cm]{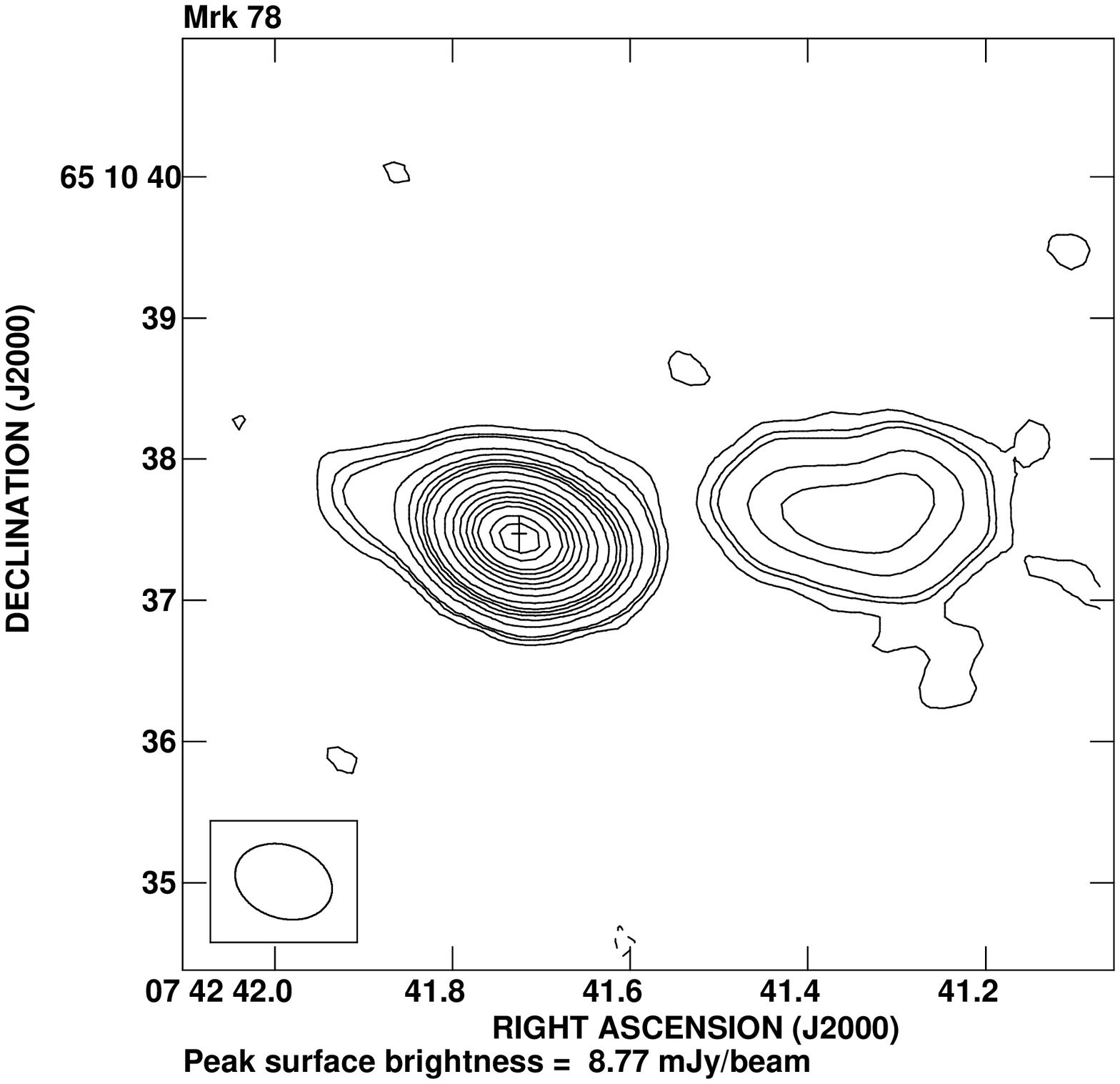} \hspace{1cm}
\includegraphics[height=6.4cm]{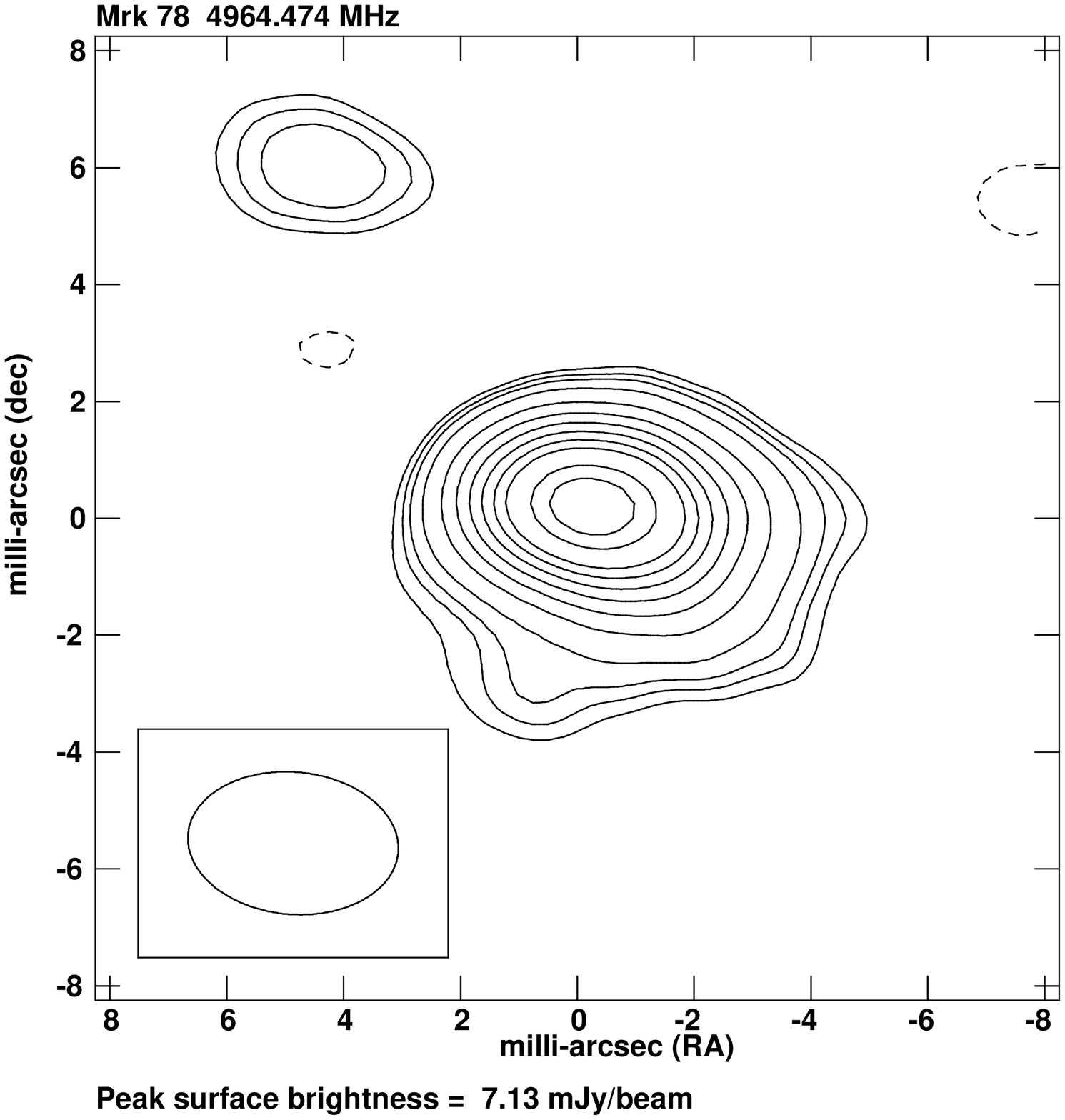} \\
\includegraphics[height=6.4cm]{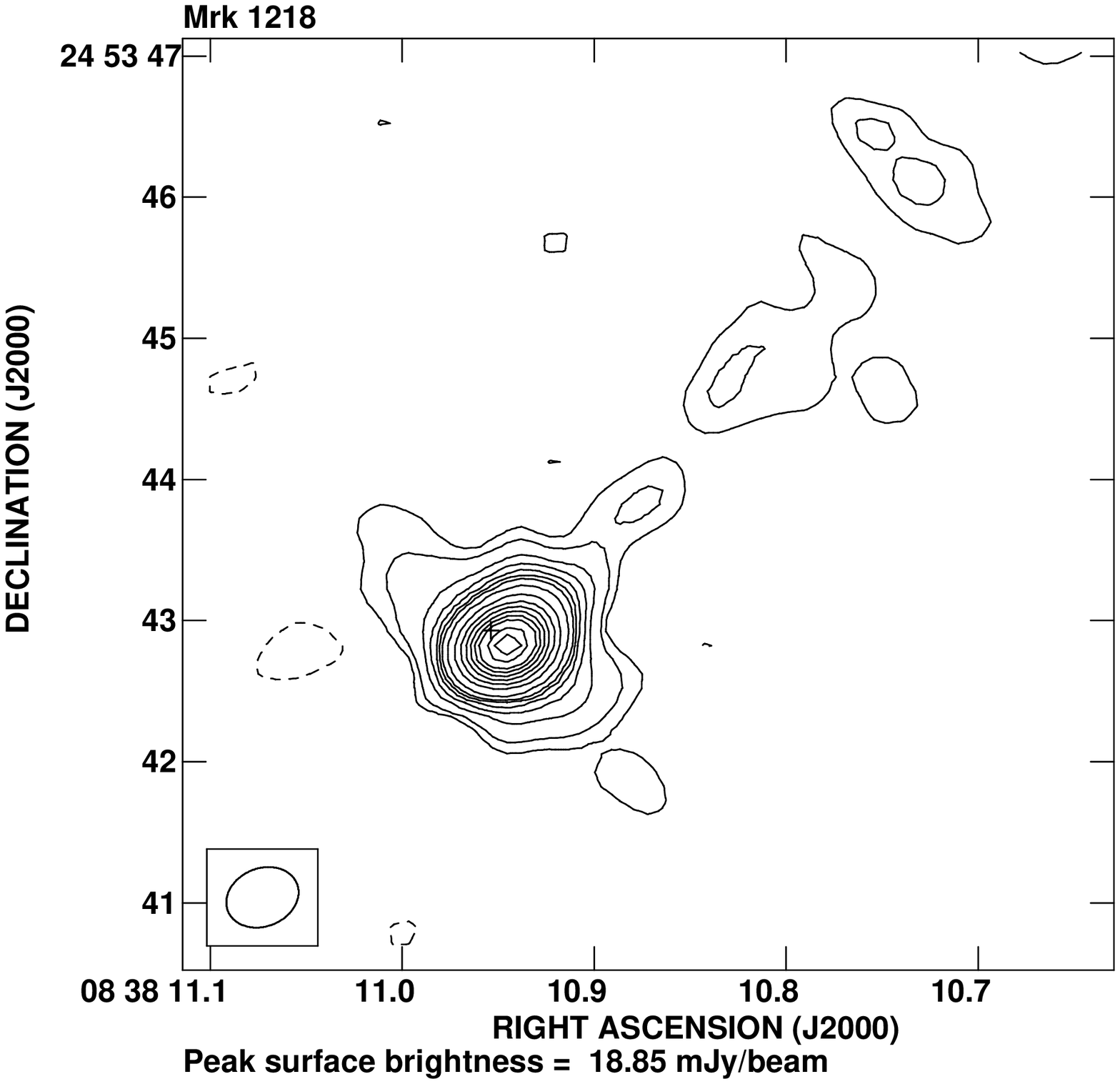} \hspace{1cm}
\includegraphics[height=6.4cm]{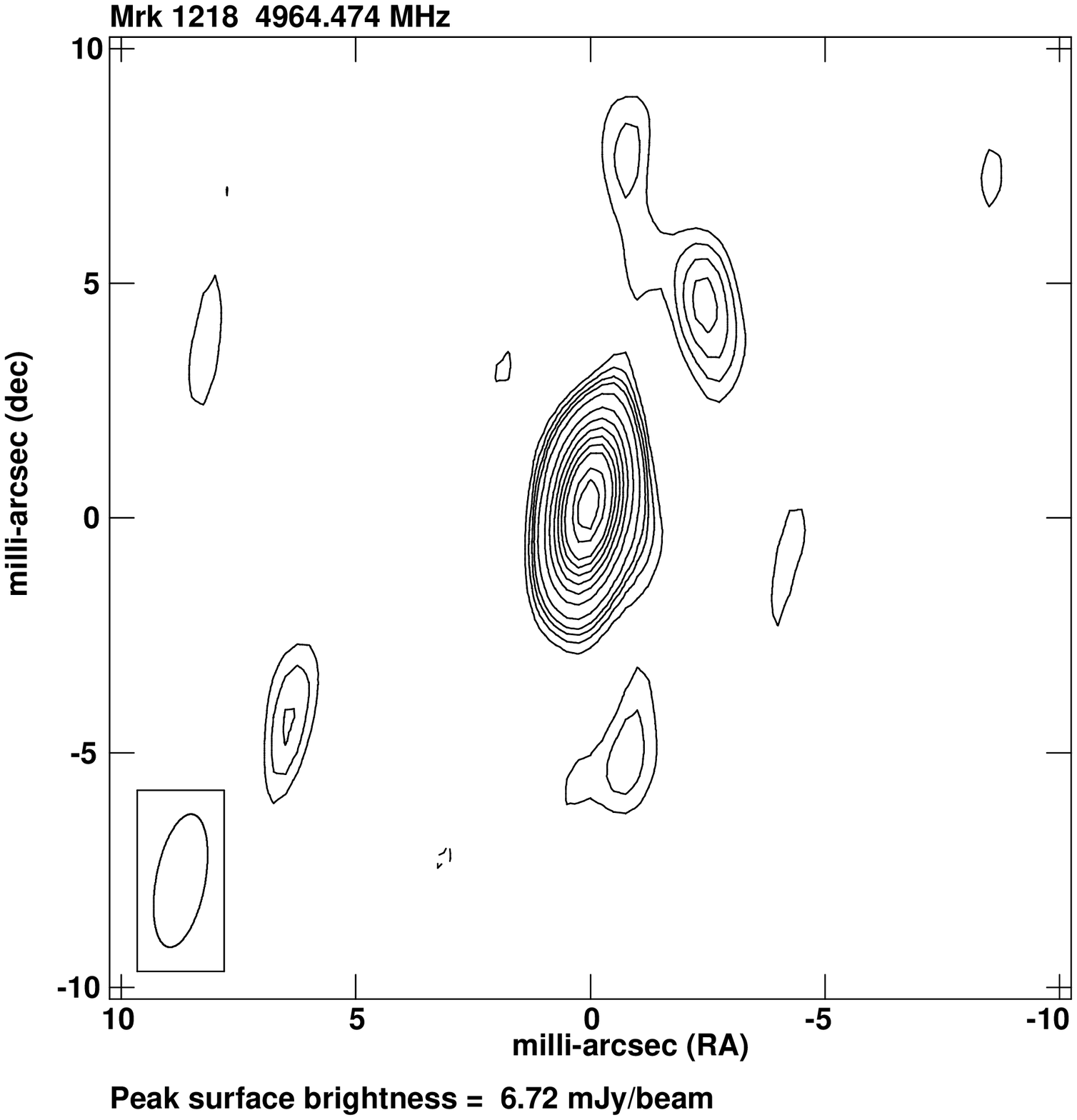} \\
\includegraphics[height=6.4cm]{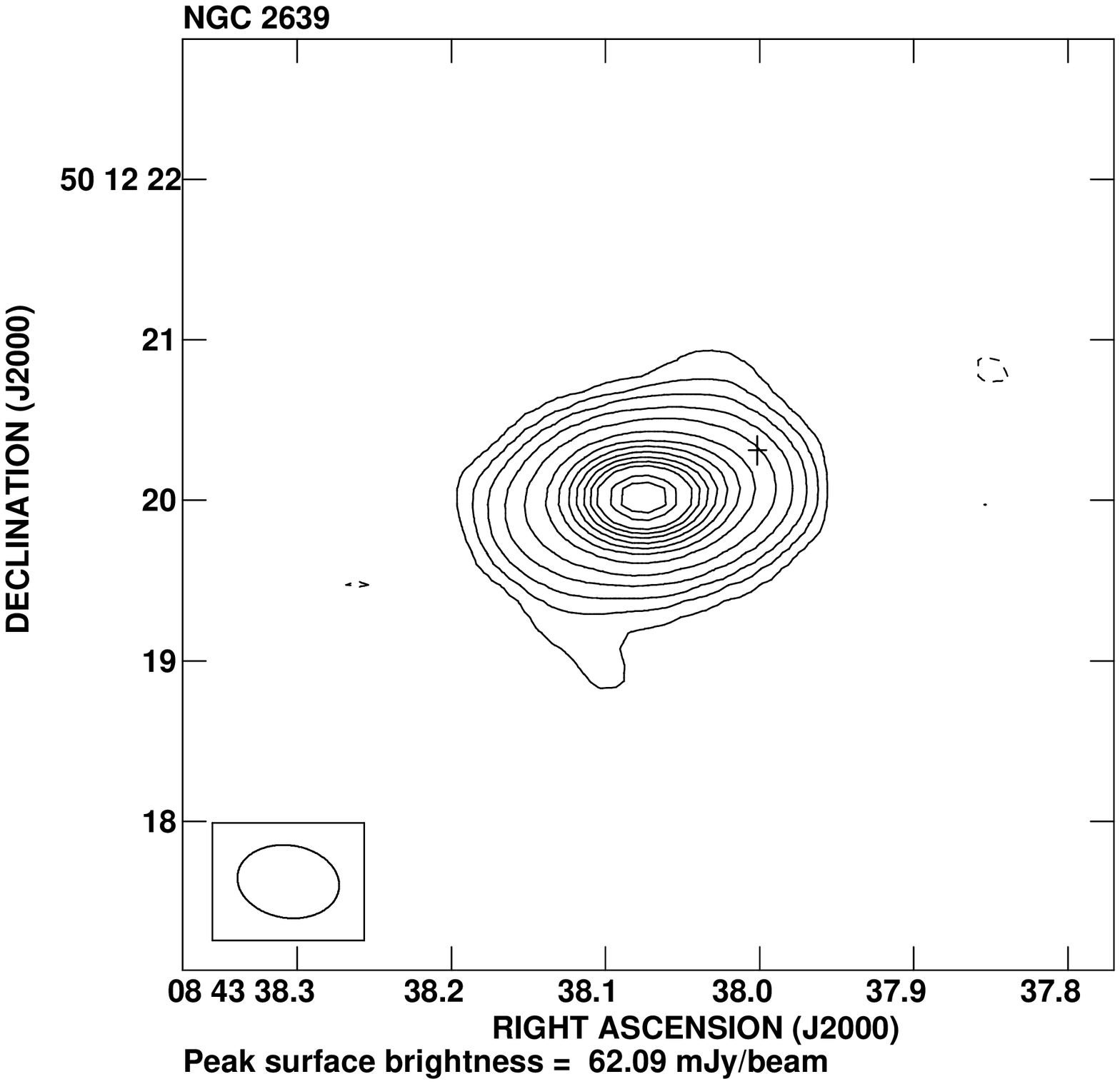} \hspace{1cm}
\includegraphics[height=6.4cm]{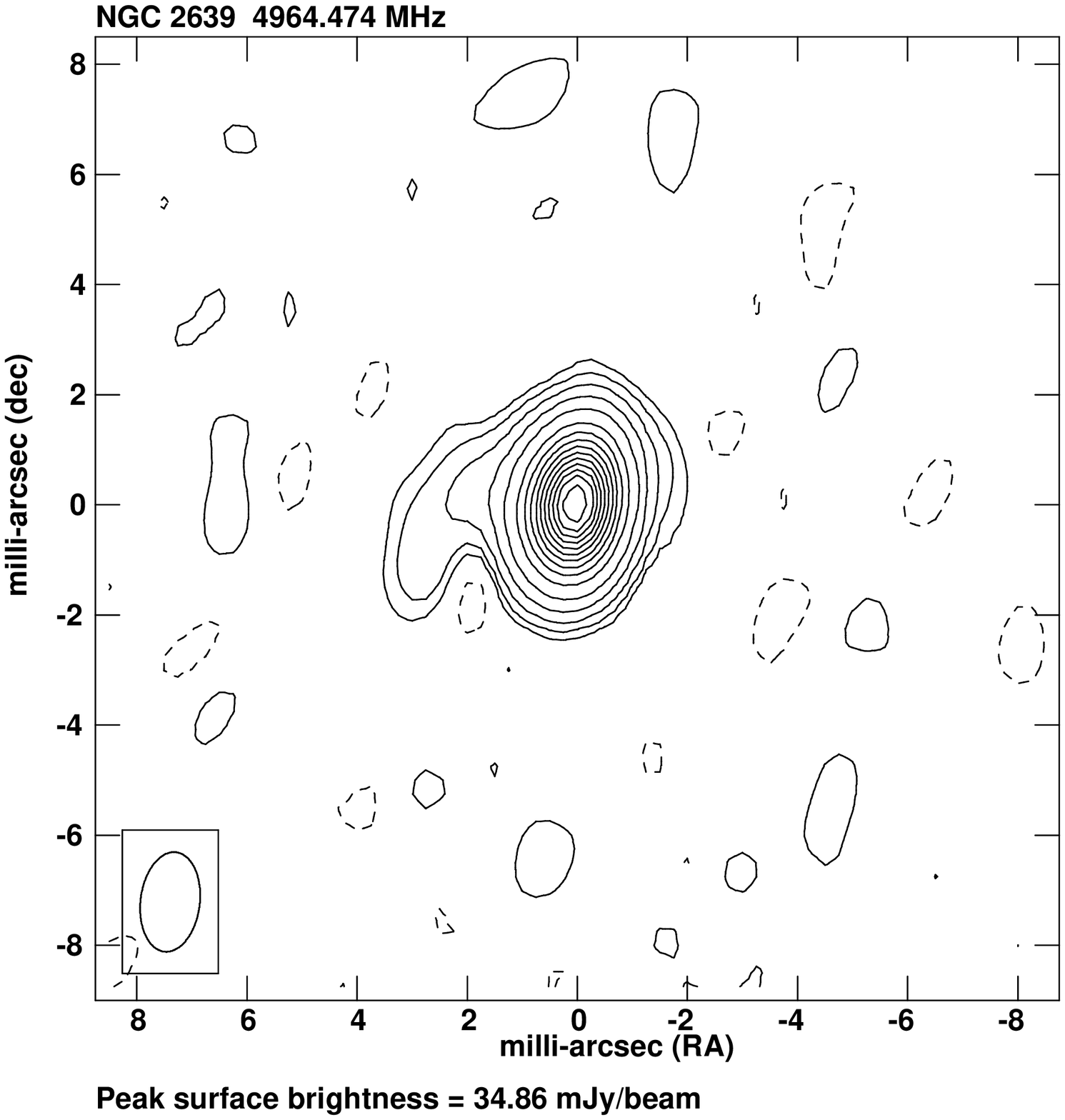} \\
\caption{({\it continued}).}  
\end{figure*}

\addtocounter{figure}{-1}
\begin{figure*}
\centering
\includegraphics[height=6.4cm]{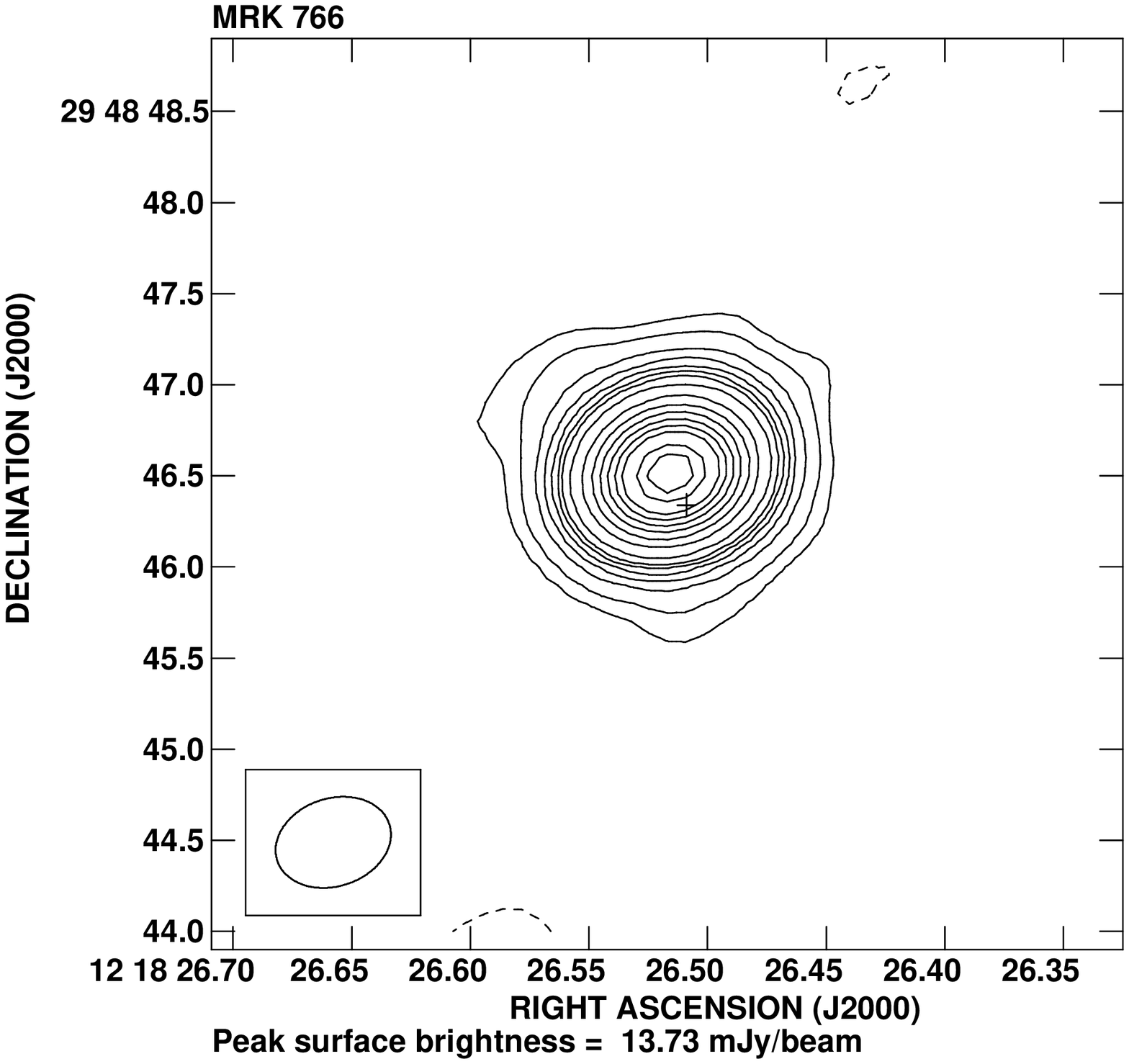} \hspace{1cm}
\includegraphics[height=6.4cm]{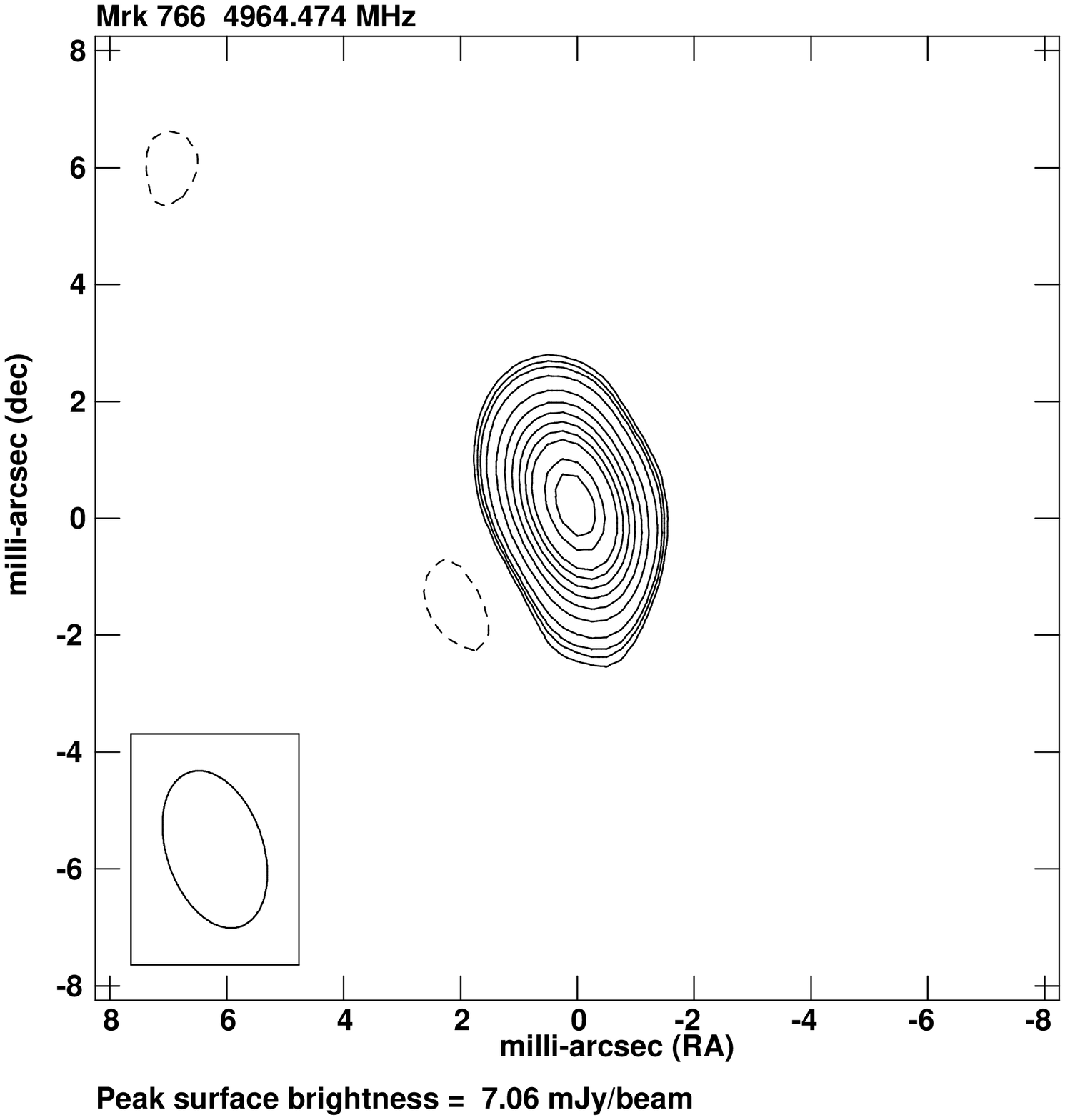} \\
\includegraphics[height=6.4cm]{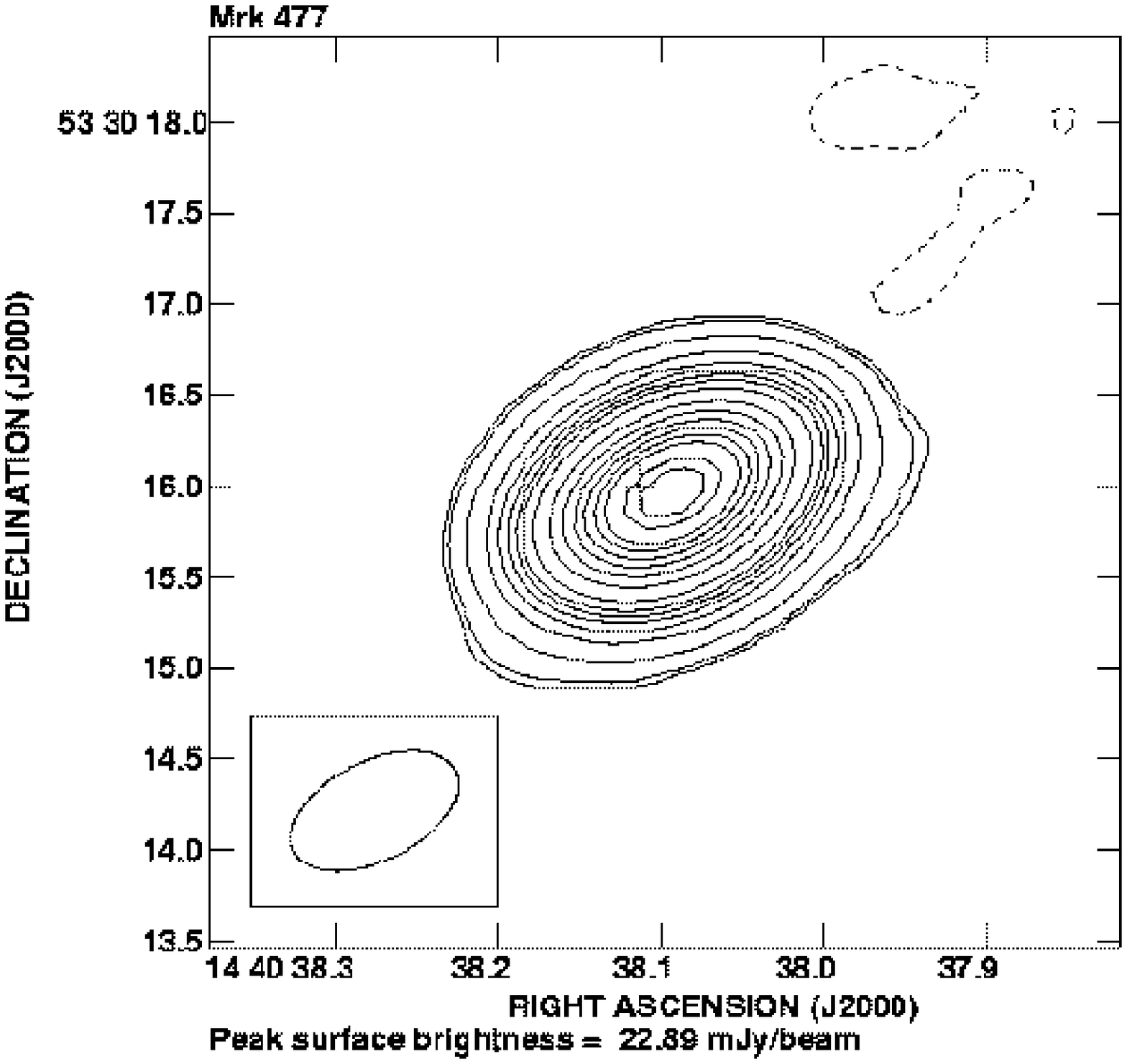} \hspace{1cm}
\includegraphics[height=6.4cm]{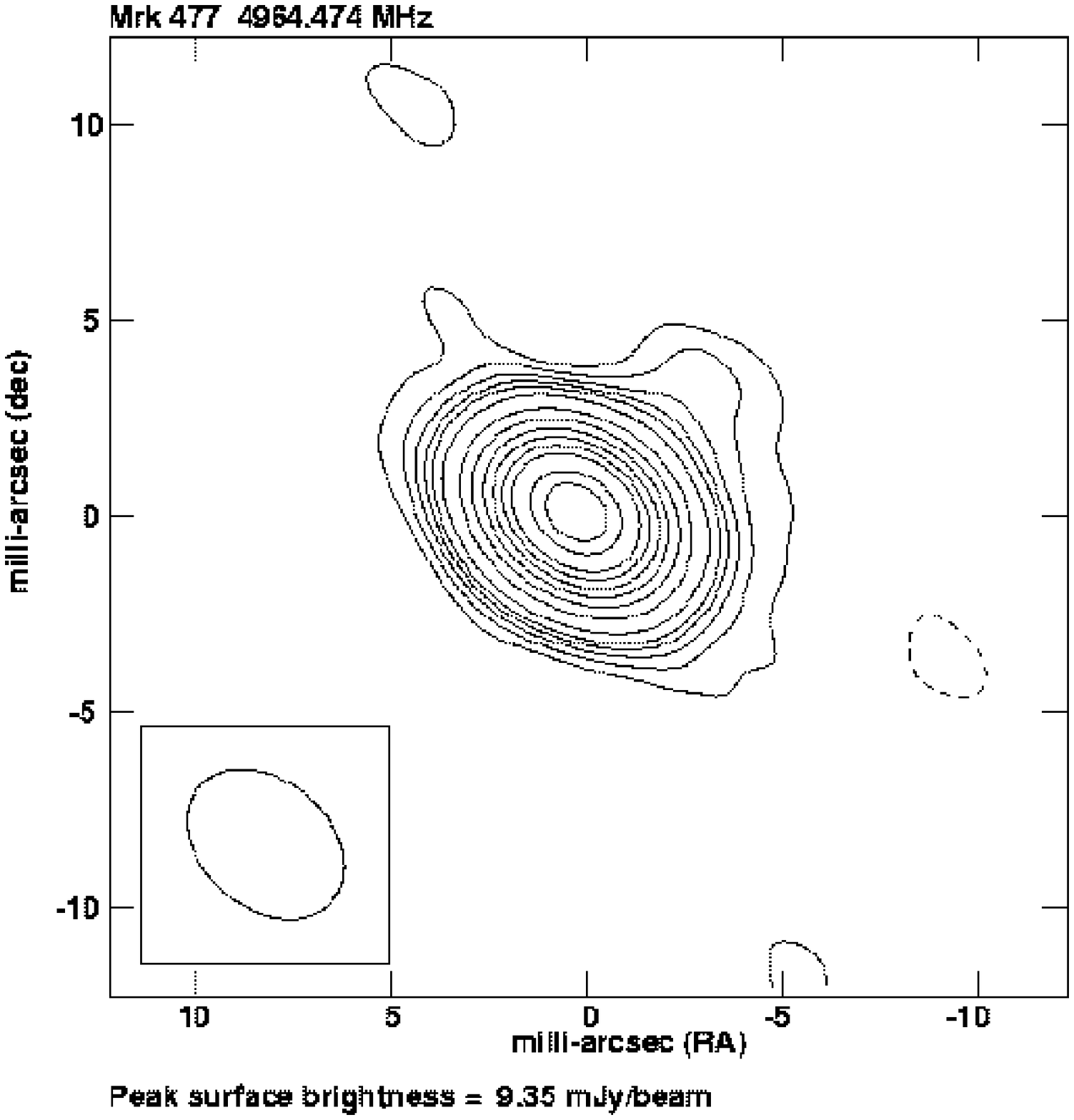} \\
\includegraphics[height=6.4cm]{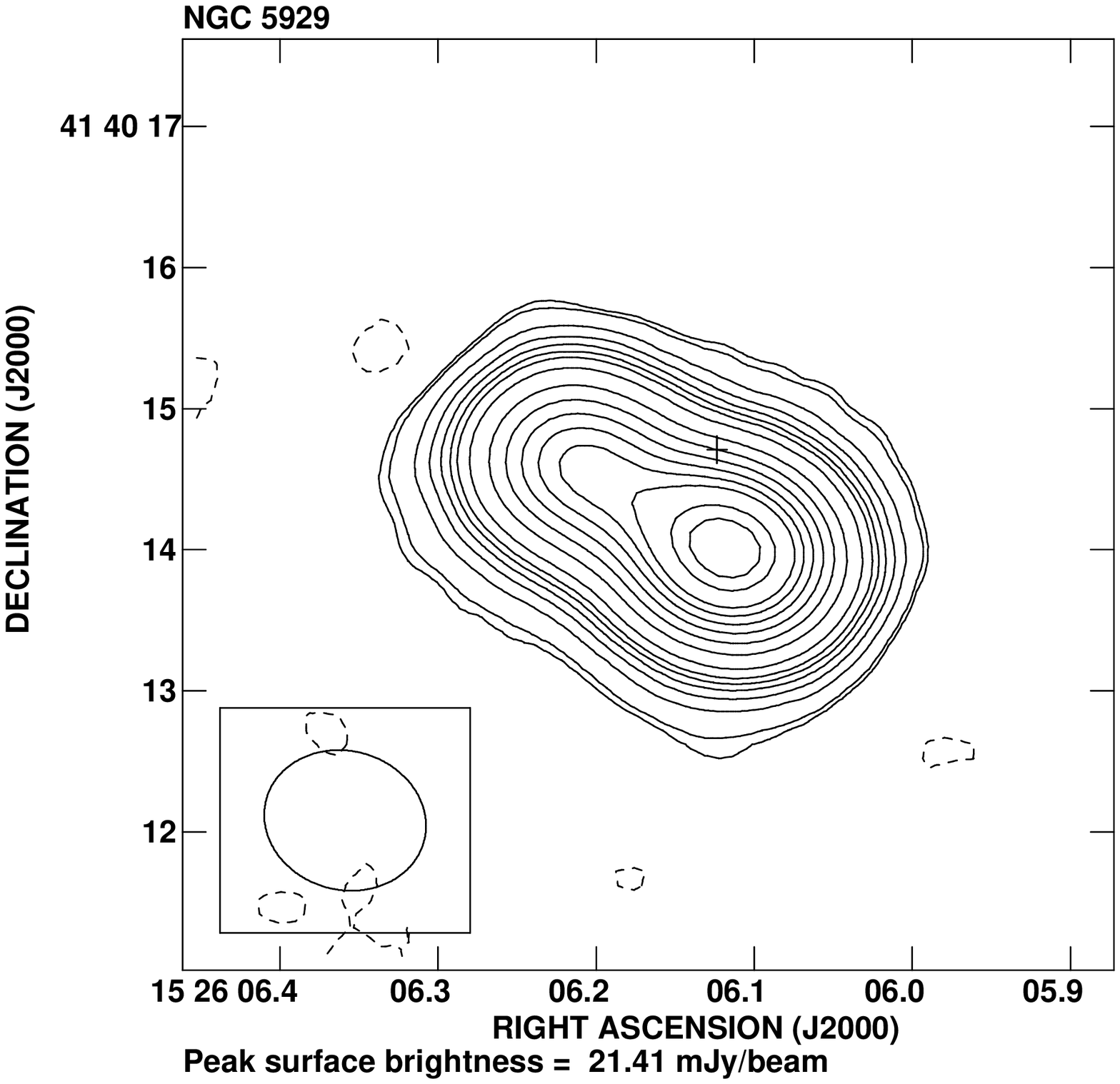} \hspace{1cm}
\includegraphics[height=6.4cm]{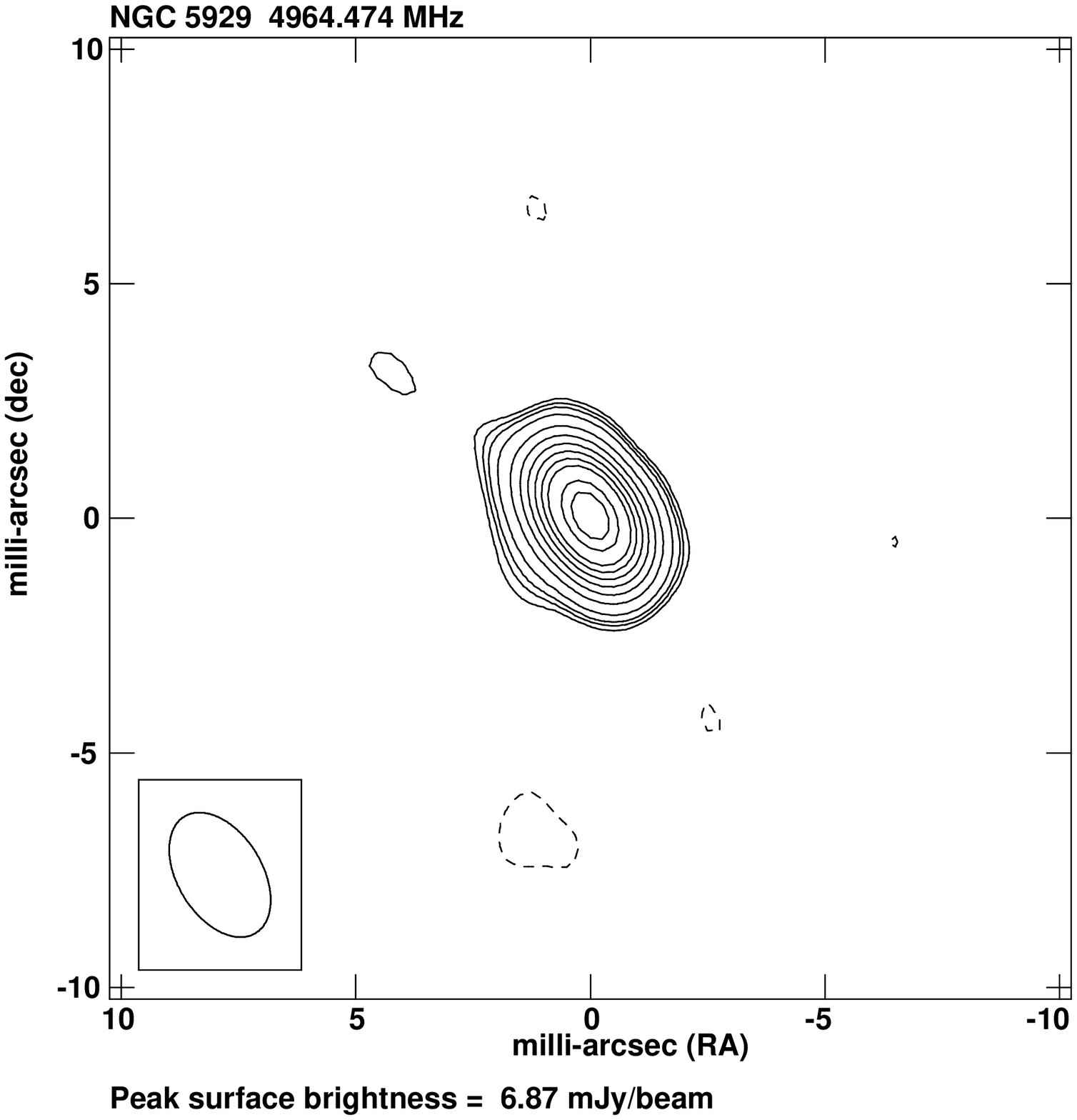} \\
\caption{({\it continued}).}
\end{figure*}

\addtocounter{figure}{-1}
\begin{figure*}
\centering
\includegraphics[height=6.4cm]{0486FY10.ps} \hspace{1cm}
\includegraphics[height=6.4cm]{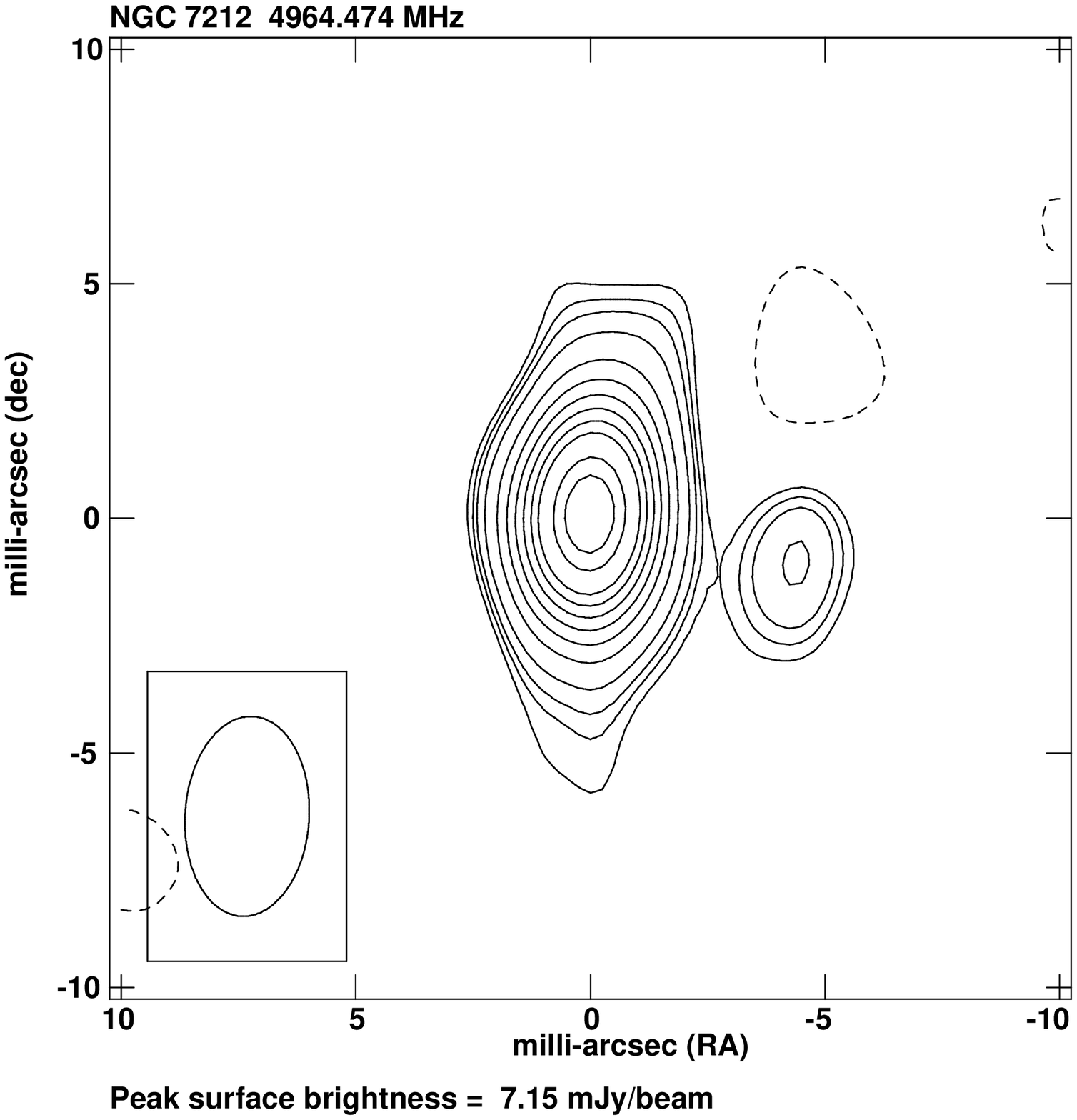} \\
\includegraphics[height=6.4cm]{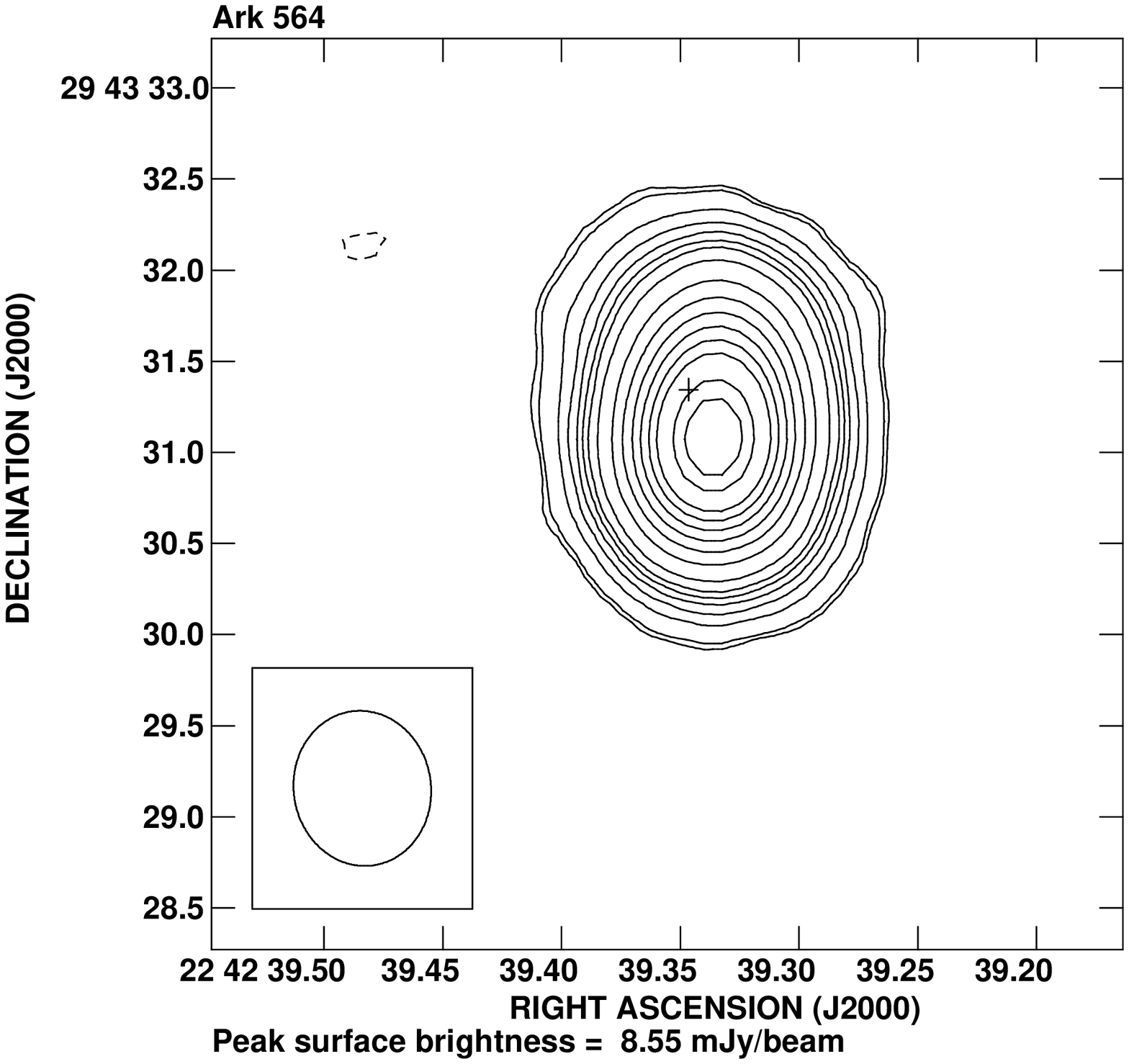} \hspace{1cm}
\includegraphics[height=6.4cm]{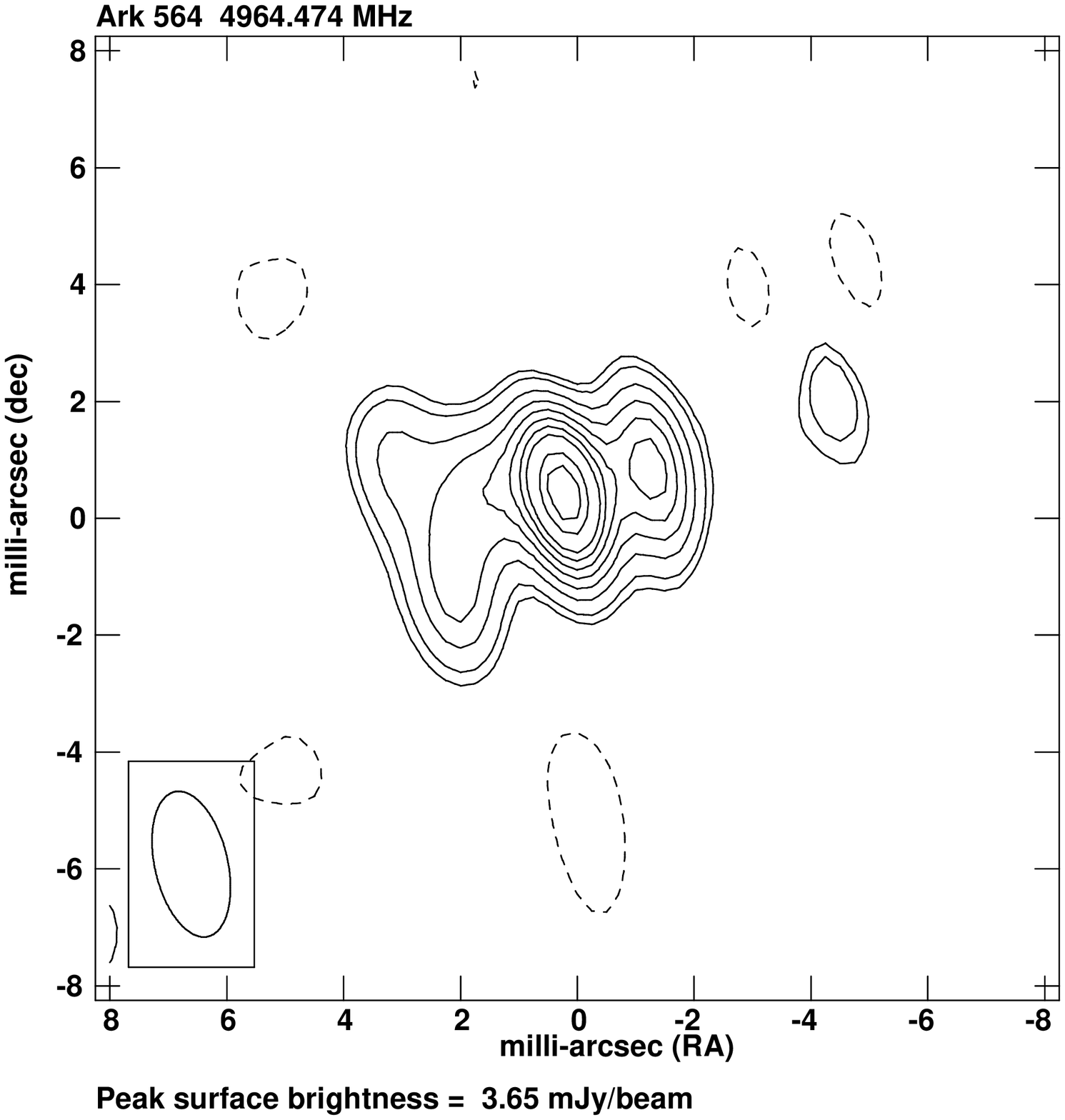} \\
\includegraphics[height=6.4cm]{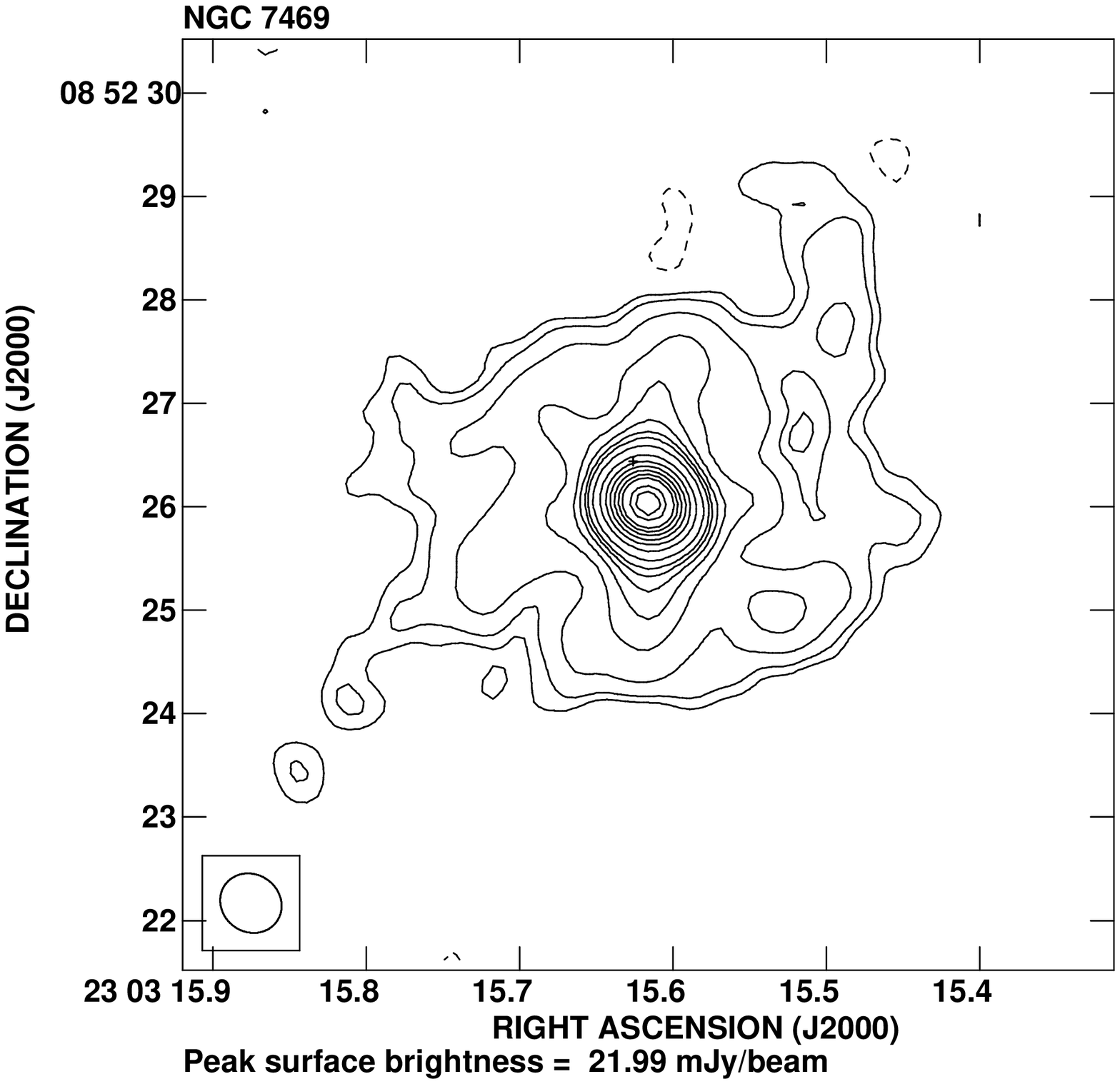} \hspace{1cm}
\includegraphics[height=6.4cm]{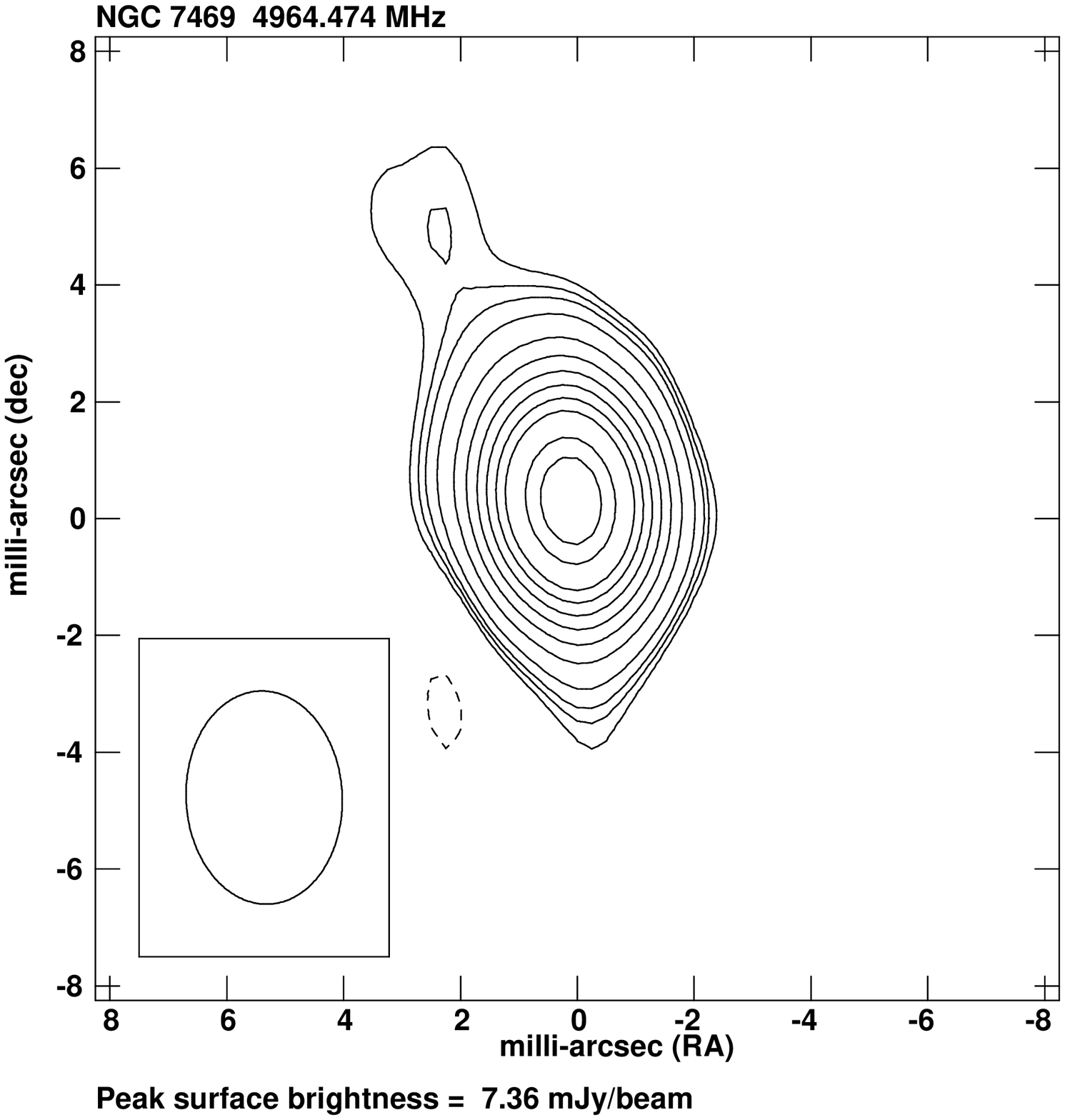} \\
\caption{({\it continued}).}
\end{figure*}

\addtocounter{figure}{-1}
\begin{figure*}
\centering
\includegraphics[height=6.4cm]{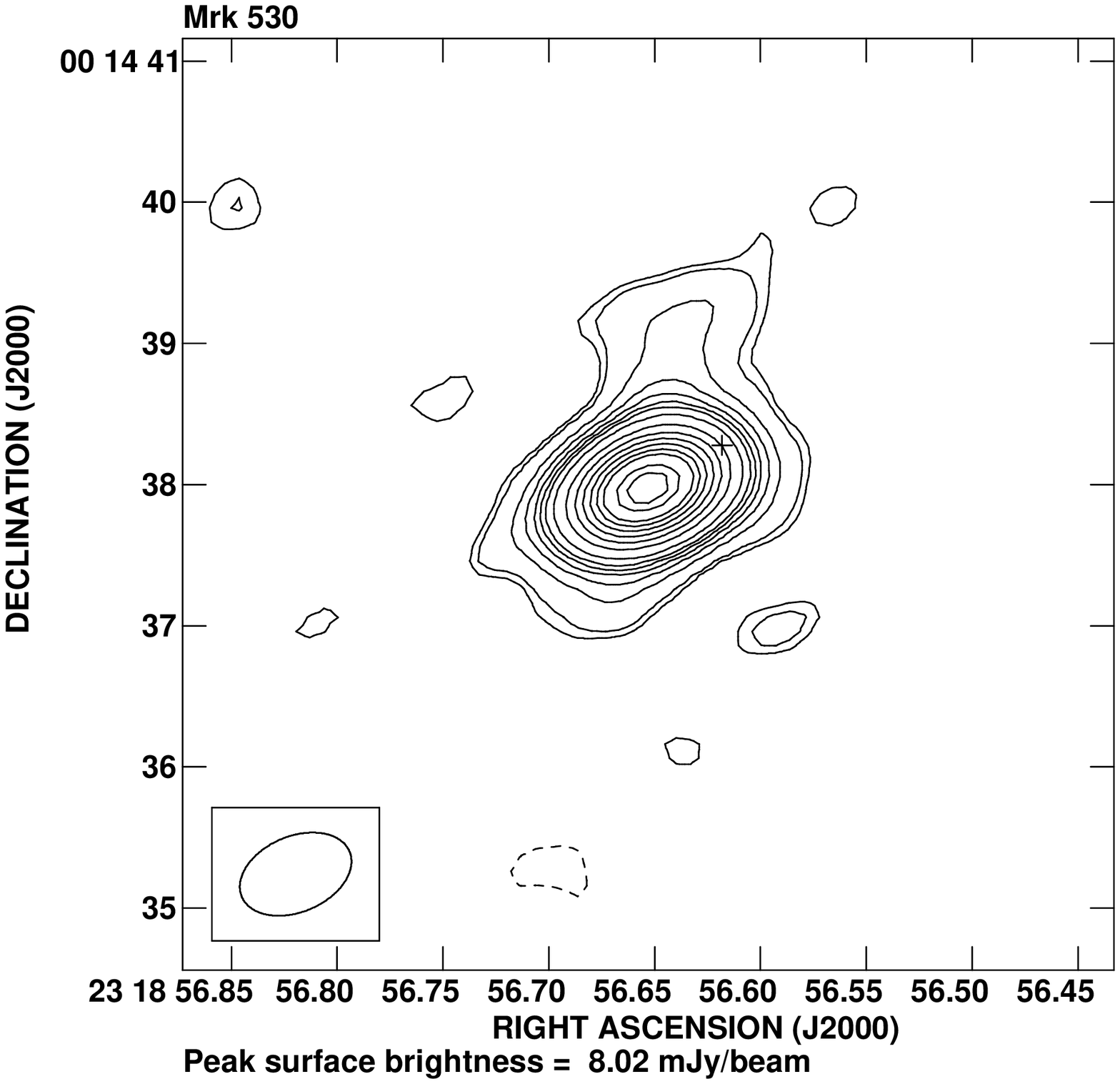} \hspace{1cm}
\includegraphics[height=6.4cm]{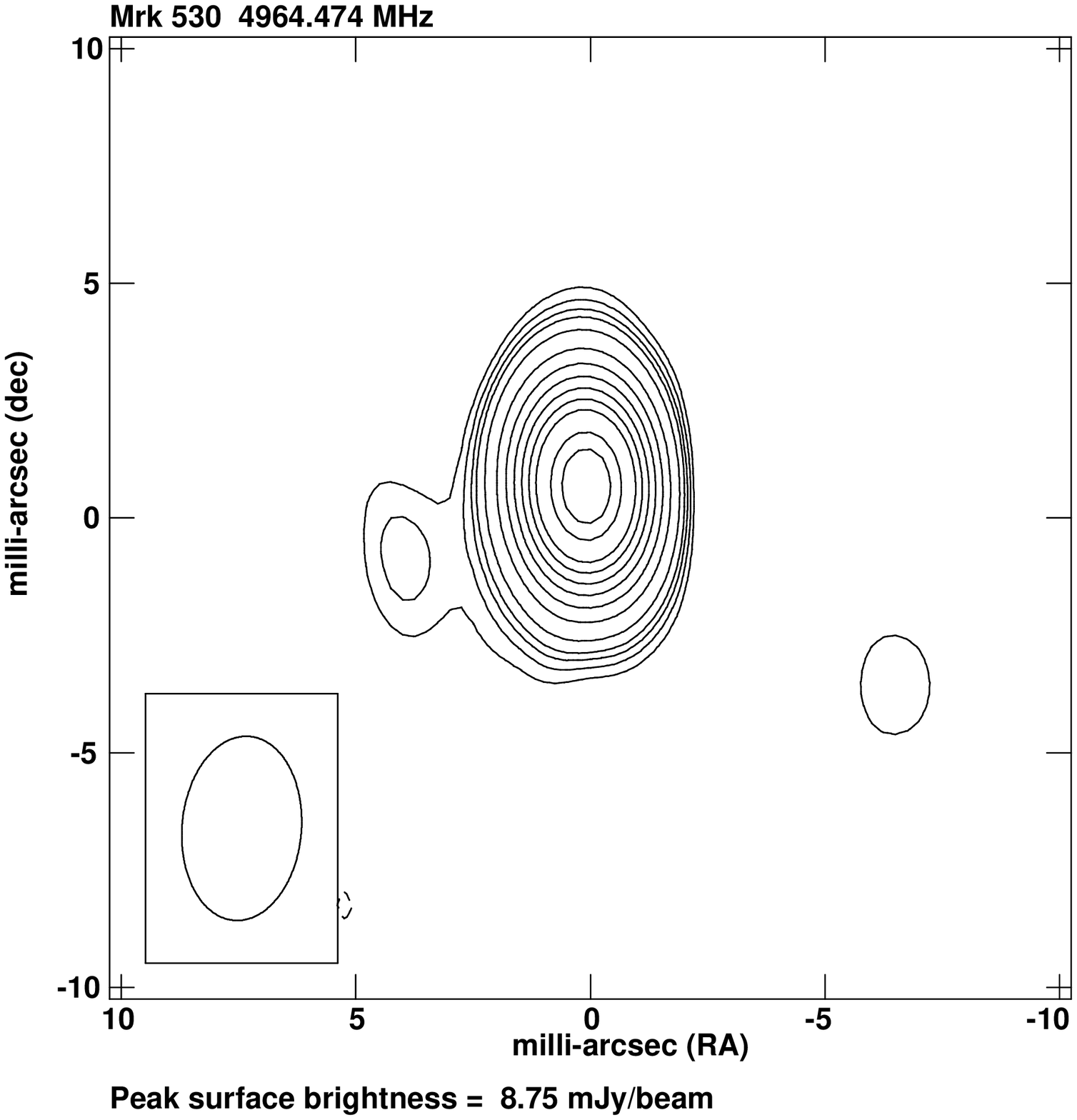} \\
\includegraphics[height=6.4cm]{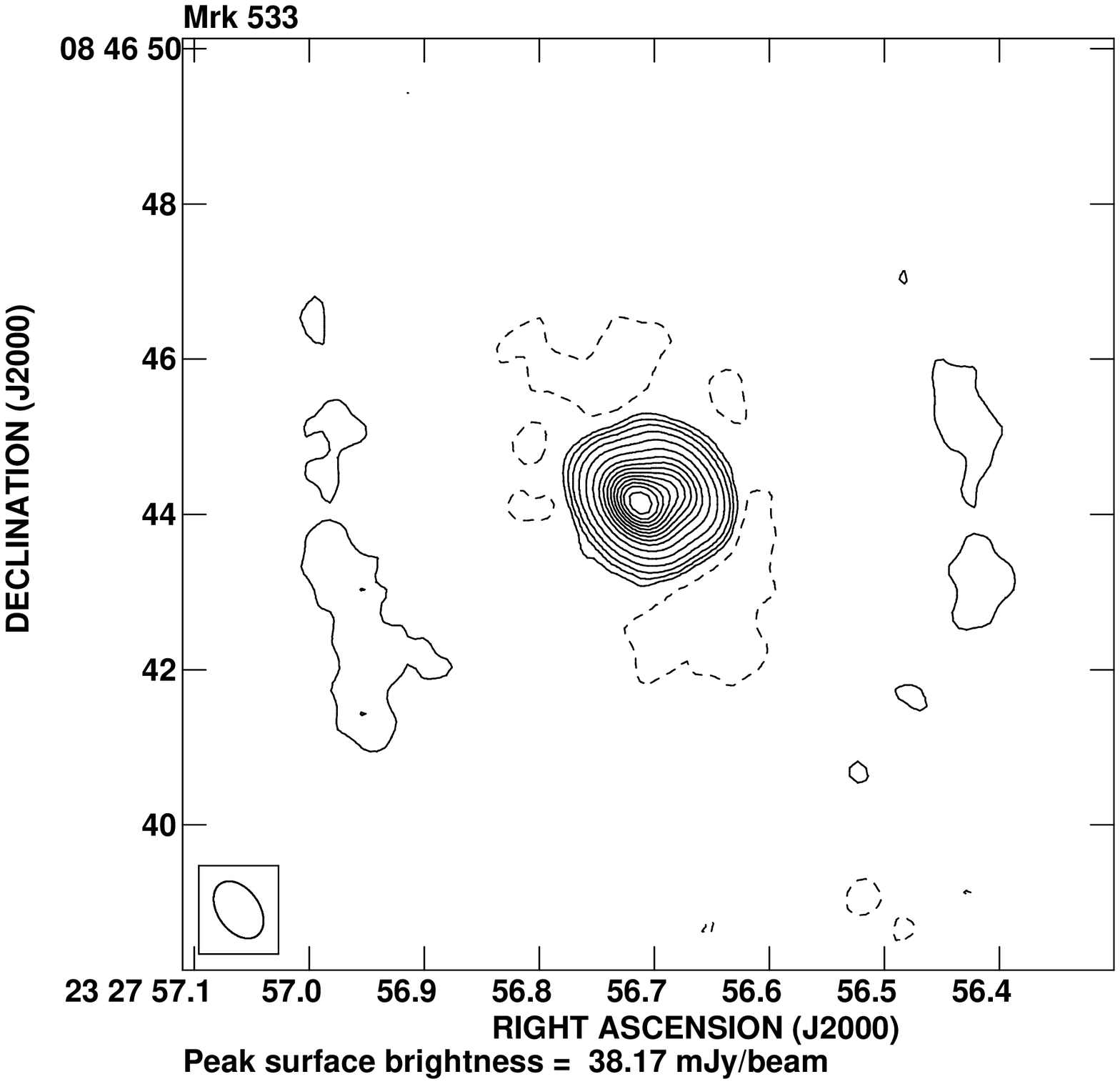} \hspace{1cm}
\includegraphics[height=6.4cm]{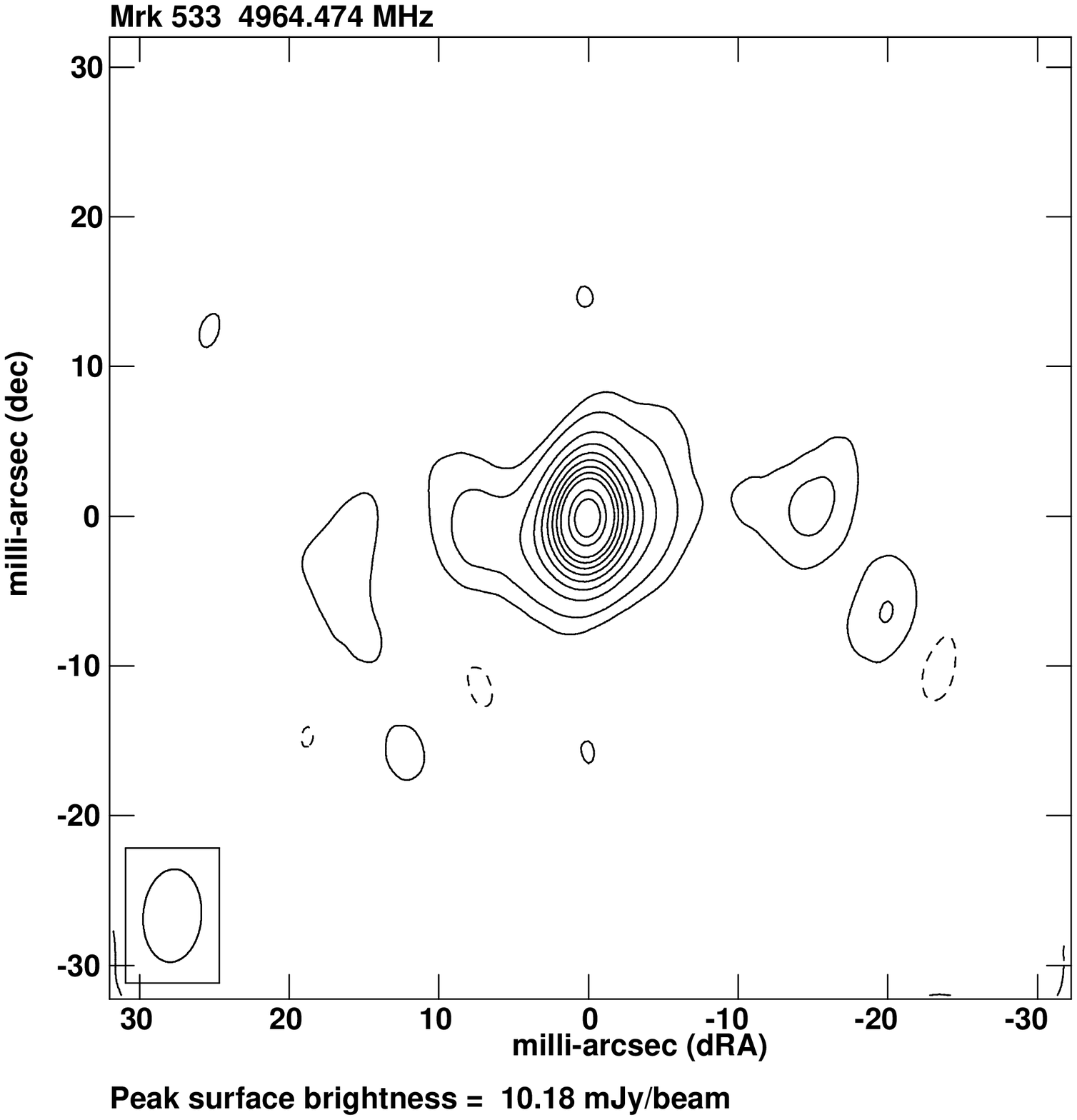} \\
\includegraphics[height=6.4cm]{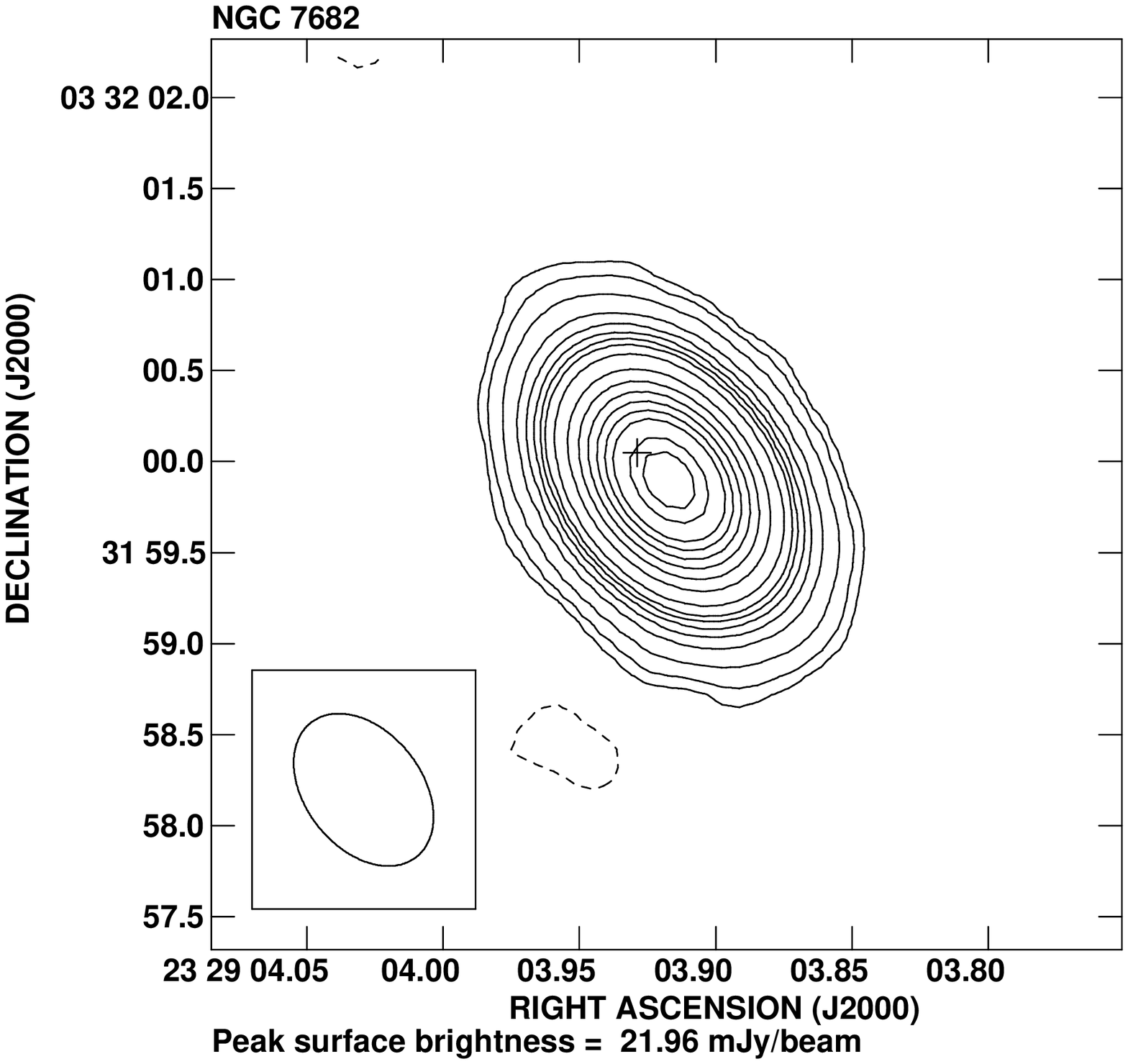} \hspace{1cm}
\includegraphics[height=6.4cm]{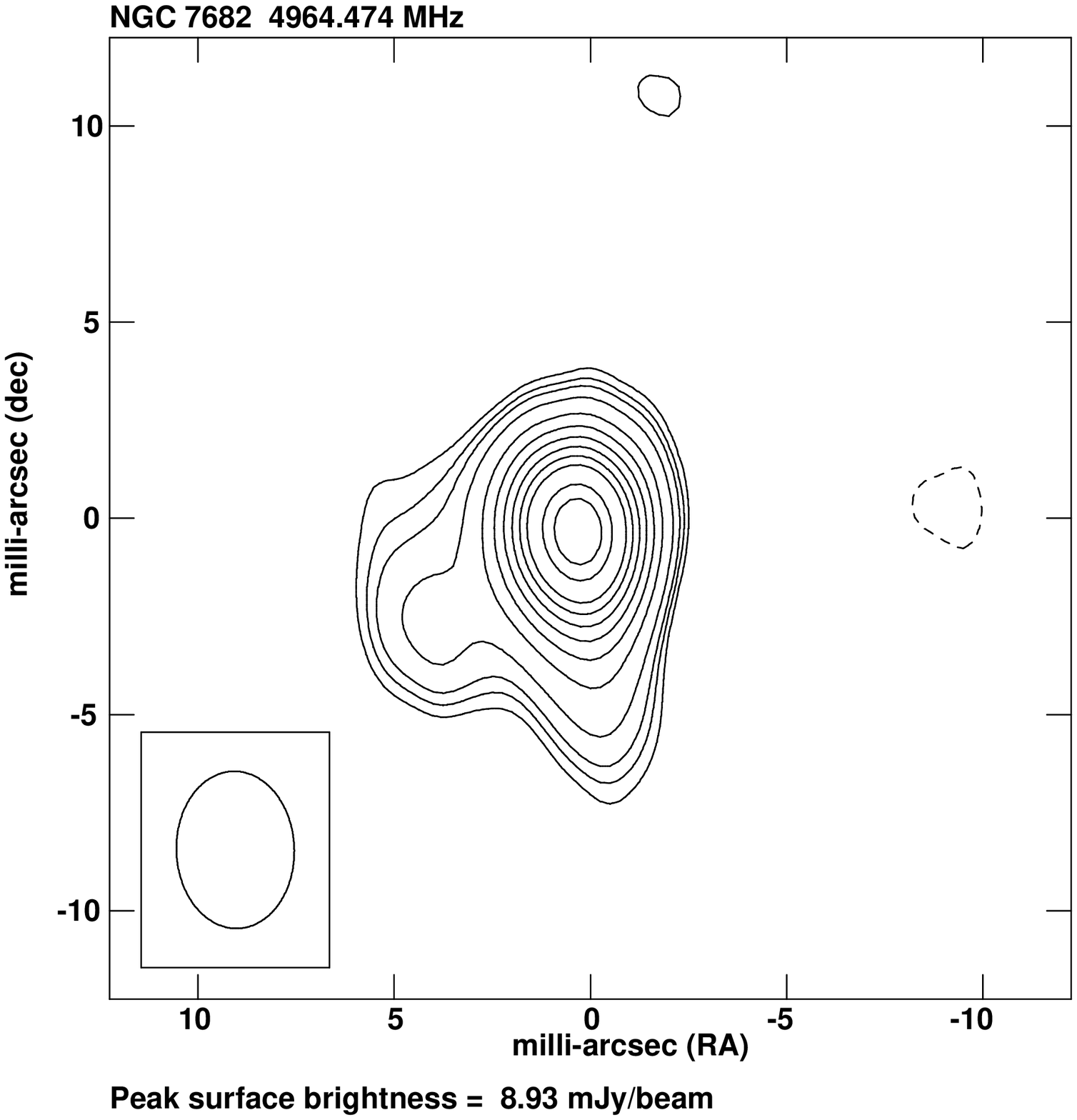} \\
\label{maps5}
\caption{({\it continued}). }
\end{figure*}

\begin{figure*}
\centering
\includegraphics[height=6.4cm]{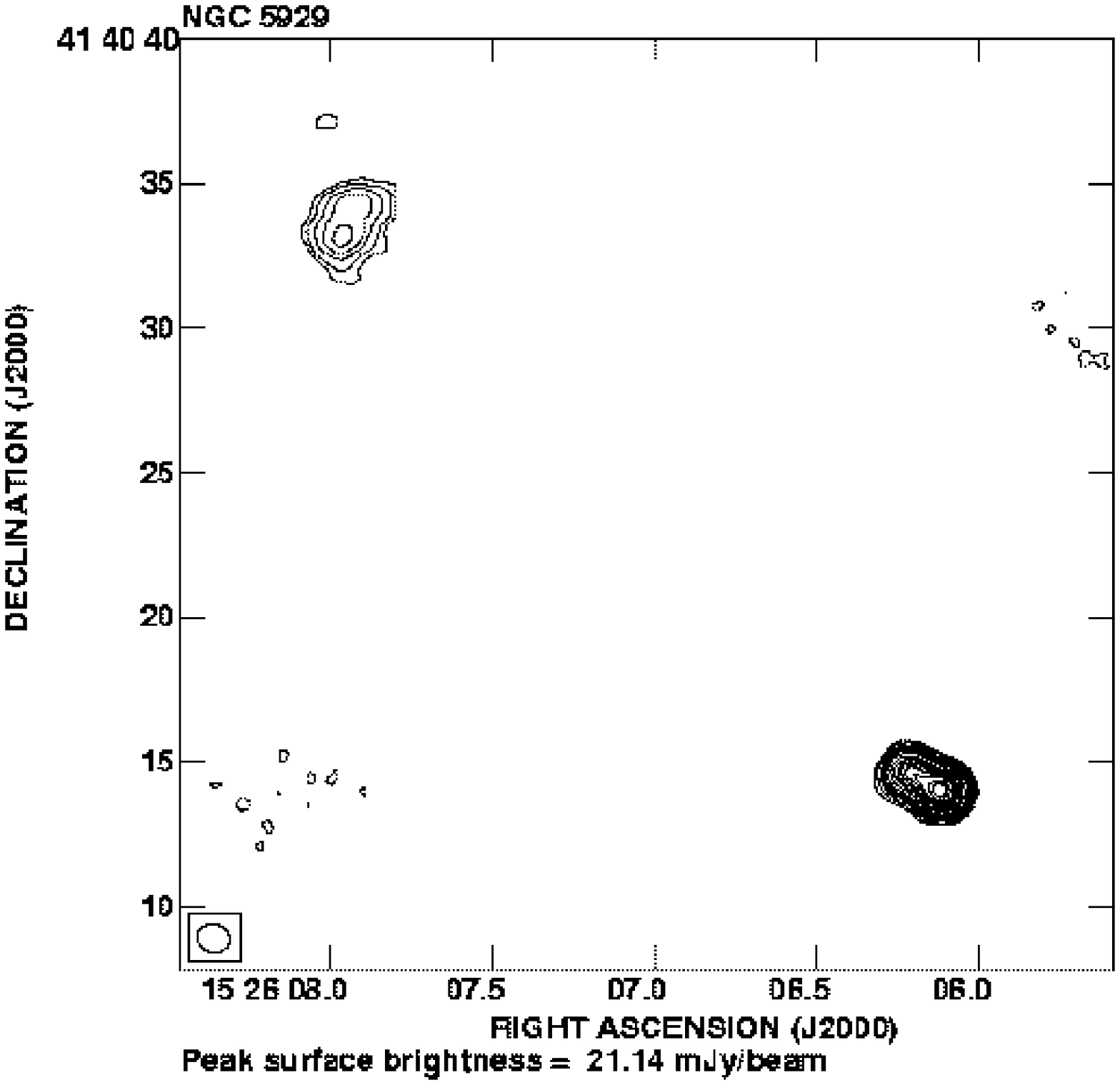} \hspace{1cm}
\includegraphics[height=6.4cm]{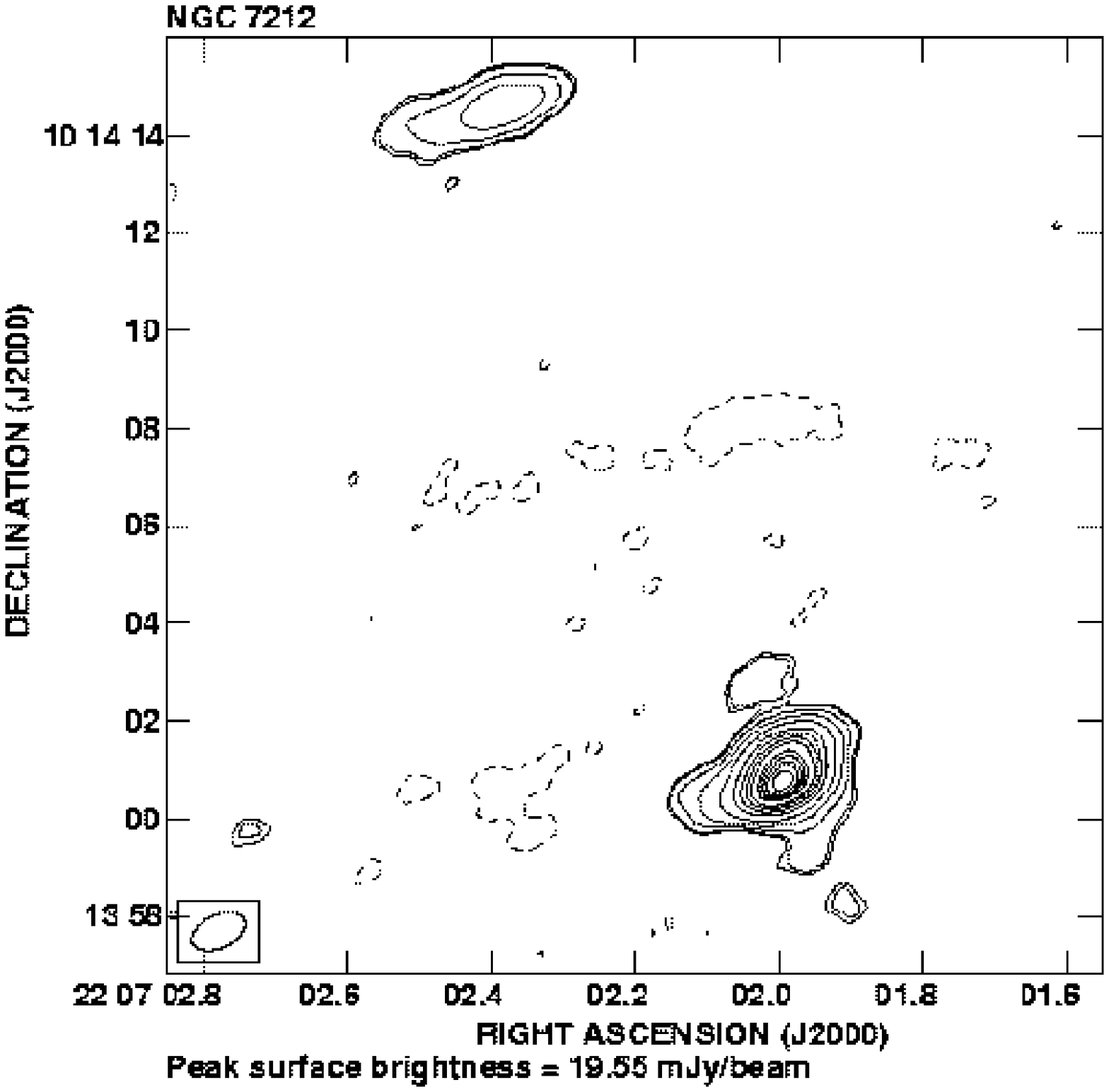}
\caption{
Left panel shows a large VLA map of NGC~5929 including
NGC~5930.
Right panel shows a large VLA map of NGC~7212 including
the emission detected from its companion to the north-east.}
\label{large_image}
\end{figure*}

\noindent {\bf NGC 2639} (0840+503)

\begin{enumerate}
\item[{\em {\small $Y$}:}]

This much studied object is a triple (P.A. of $-$77\deg) with a bright central
component \citep{UlvestadWilson89}. Our map does not
resolve the triple structure and, as in the map of \cite{UlvestadWilson89},
the radio peak lies to the south-east of the optical nucleus
\citep{ArgyleEldridge90}. This source is variable (A. Wilson, private
communication). The integrated spectral index 
obtained using the observations of
\cite{Thean00} at 8.4~GHz (July 1995/November 1996 data) 
and of \cite{UlvestadWilson89} at 5~GHz (February 1985 data), 
is $\alpha_{1.4~{\rm GHz}}^{5~{\rm GHz}}
= 0.5$. The largest angular extent of the source as measured by
\cite{UlvestadWilson89} is $\sim$~1.7~arcsec. \cite{Ho93I} classify 
this source as a LINER.

\item[{\em {\small $G$}:}]

Our map is dominated by the bright central component, with
some weaker emission to the east (P.A. $\sim$~90$^{\circ}$), as is confirmed
by our two-component model-fit. The main component contains almost
$\sim$ 80~\% of the total detected flux density. The flux density of the
bright component 31.6~mJy ($1.8\times 1.1$~mas$^2$ beam size), is lower than
the value of 47~mJy ($2.0\times 1.7$~mas$^2$ beam size) quoted
by \cite{Wilson98}, but is higher than the flux density derived from the
early VLBI measurements of \cite{Hummel82}, {\it viz.}, 27~mJy.

\end{enumerate}

\noindent {\bf Mrk 766} (1215+300)

\begin{enumerate}
\item[{\em {\small $Y$}:}]

This source is essentially unresolved in our map,
but shows hints of a weak halo-like structure. The 5~GHz
VLA~$A$ array image of \cite{UlvestadWilson84I} shows it
to be slightly resolved, with most of the extended emission
lying slightly to the north of the bright component; they derive a linear
size of $\sim$~100~pc. We detect a flux density of 15~mJy, similar to
the 5~GHz flux density found by \cite{UlvestadWilson84I}.
The radio peak and the optical nucleus \citep{Clements81} are coincident
within the errors. The spectral indices derived using our observations
along with the 8.4~GHz observations of \cite{Kukula95} (obtained 
in June 1991) and the 1.5~GHz observations of \cite{UlvestadWilson89}
(obtained in February 1985) are $\alpha_{1.5~{\rm GHz}}^{5~{\rm GHz}}$
= 0.8, and $\alpha_{5~{\rm GHz}}^{8.4~{\rm GHz}}$ = 1.1.

\item[{\em {\small $G$}:}]

Our slightly resolved image shows
a slight extension towards the south (P.A. $\sim$ 160$^{\circ}$),
which we were not able to adequately model-fit;  our model contains
about 60~\% of the total flux density on VLBI scales.

\end{enumerate}

\noindent {\bf Mrk 477} (1439+537)

\begin{enumerate}
\item[{\em {\small $Y$}:}]

Our image is essentially unresolved,
whereas the image of \cite{UlvestadWilson84I} at the same frequency
and with similar resolution was slightly resolved. The central
component in their map shows weak extensions in several different
directions. Our total flux density is consistent with theirs
within the errors, with $\sim$ 90~\% of the total flux density
being in the unresolved central component.

\item[{\em {\small $G$}:}]

The object is dominated by a compact component containing about
80~\% of the total flux density on VLBI scales, with some evidence for weak
halo-like emission.

\end{enumerate}

\noindent {\bf NGC 5929} (1524+418)

\begin{enumerate}
\item[{\em {\small $Y$}:}]

The central component of the triple structure visible in the
8.4~GHz VLA~$A$ array image of \citet{Su96}
cannot be distinguished in our lower-resolution image.
The spectral indices of the north-east and south-west components 
are steep, with $\alpha_{1.6~{\rm GHz}}^{15~{\rm GHz}} =$
0.9 and 0.8, respectively. The central component has a flatter spectrum,
with  $\alpha_{1.6~{\rm GHz}}^{15~{\rm GHz}} = 0.3$
\citep{Su96,Kukula95,WilsonKeel89} and is close to the position of the
optical nucleus, and hence is likely to be the core. 
The epochs of these observations are August 1993 (1.6~GHz), 
November 1992 (5~GHz), June 1990 (8.4~GHz) and 
March 1989 (15~GHz). We also detect
8~mJy from the unresolved neighboring galaxy NGC~5930
(Fig.~\ref{large_image}, left panel), 
as is reported by \cite{Kukula95} for VLA~$C$
configuration data at 8.4~GHz. The largest angular size of the
source is $\sim$~2.0~arcsec \citep{Kukula95}.

\item[{\em {\small $G$}:}]

Our image shows a compact source with some signs of weak extensions toward 
the east or north-east, along P.A.~$\sim$~50$^{\circ}$.
A possible weak component found in our model-fitting is located to the
north-east ($\sim$~3~mas) of the bright component. Using the VLA
core spectral index, $\alpha_{15~{\rm GHz}}^{1.6~{\rm GHz}}$
$\simeq$ 0.32 \citep{Su96}, the core should have flux density of 1.1~mJy at
5.0~GHz. We detect a total flux density of 6.8~mJy in our
VLBI image, and therefore conclude that the detected mas-scale
component corresponds to one of the two steep spectrum components of the
triple \citep{Su96}, and not to the core. We have tried to map a larger
field to image the second steep-spectrum component, but were not able to
detect it. It is possible that this second component is resolved out by
our VLBI data; however, given that the spacing between the two
components is $\sim$~1.3~arcsec \citep{Su96}, it is likely that the variations
in the visibilities would be too rapid to be tracked with the $(u,v)$ and
time coverage of our data. Thus, we feel that our data do not rule out
the presence of a second compact component coincident with the distant
steep-spectrum lobe.

\end{enumerate}

\noindent {\bf NGC 7212} (2204+095)

\begin{enumerate}
\item[{\em {\small $Y$}:}]

Our map does not resolve the compact double separated by
0.7~arcsec to the north-west as clearly as the 8.4~GHz,
VLA~$A$ array map of \cite{Falcke98}. However, our map does show
an obvious extension towards the north (P.A. $\sim$ 10$^{\circ}$), as well
as fainter extensions roughly to the south and south-east.
The largest angular size of the source is $\sim$~3~arcsec.
We note that the position of the phase calibrator that we used was
in error at the time of our observations,
but has since been updated by the NRAO; we have incorporated this correction
in our image using the AIPS task OGEOM. NGC~7212 is 
in an interacting system \citep{Wasilewski81} and we detect faint radio 
emission from its closest companion, seen \abt~14$^{\prime\prime}$ to
its north-east (Fig.~\ref{large_image}, right panel).
The companion is visible in the Digital Sky Survey images and is clearly 
delineated in the 2MASS images. 

\item[{\em {\small $G$}:}]

Apart from the bright primary component, we detect a secondary
nearly to the west (P.A. $\sim$~$-$110$^{\circ}$), as well as
an extention toward the north, roughly in the direction of the
arcsecond-scale structure. About 90~\% of the total VLBI-scale flux density
is contained in our two-component model.

\end{enumerate}

\noindent {\bf Ark 564} (2240+294)

\begin{enumerate}
\item[{\em {\small $Y$}:}]

Although our map does not clearly resolve the triple
structure extending to 320~pc in P.A. = 6$^{\circ}$ seen by
\cite{Schmitt01} (8.4~GHz, VLA~$A$ array),
it is extended in the direction of this structure.
Our radio peak coincides with the optical nucleus
\citep{Clements81} within the errors. The total flux density
at 5~GHz is the same as that quoted by \cite{Ulvestad81}
within the errors.

\item[{\em {\small $G$}:}]

This is the faintest source in our sample and was detected on 
the least number of baselines. Accordingly, the $(u,v)$ coverage
obtained was relatively poor. Although our map shows an elongated 
structure with four components in P.A. $\sim$ 110$^{\circ}$, we do not feel
confident that this is an accurate representation of the VLBI-scale
structure of the source. Attempts to model-fit this structure were
not successful, and so our single-component fit has essentially fit 
the parameters of the brightest component. The inadequacy of the fit
as a description of the VLBI-scale emission as a whole is reflected
in the fact that only 35~\% of the total VLBI-scale flux density is
reproduced by the model.

\end{enumerate}

\noindent {\bf NGC 7469} (2300+086)

\begin{enumerate}
\item[{\em {\small $Y$}:}]

Our image shows a halo $\sim$~4~kpc in extent around the dominant compact
component, which is coincident with the optical nucleus \citep{Clements81},
similar to the images in \cite{Wilson91}. 
\cite{Wilson91} determine the spectral index to be
$\alpha_{5~{\rm GHz}}^{15~{\rm GHz}}$ \abt ~0.9. The features seen in
this source are also seen by \cite{Condon82} and \cite{Ulvestad81} in
their lower resolution images. The 8.4~GHz map presented by
\cite{Kukula95} also shows a point source surrounded by faint emission.
Using our flux density and that given by \cite{Kukula95} at
8.4~GHz (obtained in June 1991), we obtain the integrated spectral index
$\alpha_{5~{\rm GHz}}^{8.4~{\rm GHz}}$ = 0.9, similar to the
$\alpha_{5~{\rm GHz}}^{15~{\rm GHz}}$ value determined by
\cite{Wilson91}.

\item[{\em {\small $G$}:}]

Most of the detected emission is from a  compact component, though
there appears to be an extension toward the north-east.
This is consistent with the prediction of \cite{Smith98II},
who, based on their observed visibility functions, suggested the presence
of multiple components. Using our detected flux density
and the 1.4~GHz correlated flux density measured by
\cite{Smith98II} of 12~mJy (September 1991), we obtain the spectral index
$\alpha_{1.4~{\rm GHz}}^{5~{\rm GHz}}$ = 0.5.

\end{enumerate}

\noindent {\bf Mrk 530} (2316$-$000)

\begin{enumerate}
\item[{\em {\small $Y$}:}]

This source was reported to be a compact point source by  \cite{Kukula95}
at 8.4~GHz, whereas our 5~GHz map
shows a compact source and a faint
(3~$\sigma$) extension to the north (P.A. $\sim$ $-$10$^{\circ}$).
Comparing our total flux density with the 3.26~mJy found
by \cite{Kukula95} at 8.4~GHz in June 1991, we derive the
spectral index $\alpha_{5~{\rm GHz}}^{8.4~{\rm GHz}}$ = 2. Our total
flux density is consistent with the measurements of \cite{Edelson87}
(11.5~mJy), and \cite{Roy94}, \viz a correlated flux density of 11~mJy
obtained using a single 275~km baseline at 2.3~GHz.

\item[{\em {\small $G$}:}]

Our model-fitting suggests that, in addition to the faint extension
to the south-east seen in the map, the compact core also has
an extension \abt~1.9~mas to the south-west.  The total flux density is
similar to the flux density associated with the central component in the
arcsec-scale image. This implies that almost all of the radio emission
detected on arcsec-scale image is associated with this compact component.

\end{enumerate}

\noindent {\bf Mrk 533} (2325+085)

\begin{enumerate}
\item[{\em {\small $Y$}:}]

Our image is consistent with the image of \cite{Momjian03}, the
 double structure seen
in the 8.4~GHz VLA~$A$ configuration map of \cite{Kukula95}
and the 1.6~GHz MERLIN map of \cite{Unger86}. The
structural P.A. is $\sim$ $-$60$^\circ$. The peak of the brightest 
radio component (component `C' of \cite{Momjian03}) is $\sim$~0.4~arcsec
to the south of the optical position of \cite{Clements83}.

\item[{\em {\small $G$}:}]

On mas-scales, we detect only component `C' of \cite{Momjian03}, 
which forms a linear radio source extending in the east-west direction. 
Our model-fitting indicates the
presence of five components. The CLEAN components are distributed
out to $\sim$ 15~mas on either side of the phase centre (total extent
of $\sim$ 30~mas) along P.A. $\sim$ $-$80$^{\circ}$, similar to the P.A.
of the arcsec-scale radio structure ($\sim$~$-$60$^\circ$).
 Our model contains about 80~\% of the total flux density 
that we detect on VLBI scales.

\end{enumerate}

\noindent {\bf NGC~7682} (2326+032)

\begin{enumerate}
\item[{\em {\small $Y$}:}]

Our map is unresolved, as was also true of the image
of \cite{Kukula95}. Combining our flux density with the 8.4~GHz
flux density given by \cite{Kukula95} (obtained in June 1991)  yields a spectral
index $\alpha_{5~{\rm GHz}}^{8.4~{\rm GHz}}\simeq~1.1$. The radio peak
coincides with the optical nucleus \citep{Clements83} within the errors.

\item[{\em {\small $G$}:}]

The image shows  clear evidence for extended structure to the south 
(P.A. $\sim$ 180$^{\circ}$) and south-east (P.A. $\sim$ 120$^{\circ}$). The best
model-fit to the visibilities was a combination of one circular Gaussian and 
two point sources which accounts for most (about 80~\%) of this emission.

\end{enumerate}

\section{Conclusions}
\label{conclu}

We have detected all the 15 Seyferts that we observed with VLBI.
We have presented here the VLBI images and simultaneous
VLA images that we obtained.

In \abt~30~\% of the cases, we detect
a single component on mas-scales.  In 40~\% of the cases, at least half
of the total emission on arcsec scales is detected in the VLBI
images. In Mrk~530, almost all the emission from the arcsec-scale core
is present on mas-scales. The measurements presented here
represent most of the database that we require for rigorously testing
the predictions of the US for Seyferts with respect to their
parsec-scale radio structure. We will present the results of such tests
in an accompanying paper.

\begin{acknowledgements}
Our many thanks to the support staff of the European VLBI Network
and the National Radio Astronomy Observatory,
and especially to Joan Wrobel and Craig Walker for help in
scheduling the VLBI observations.
We also thank the referee, Dr. Alan Roy, for his prompt review
of the manuscript, and for useful and detailed comments that
lead to improvement of the paper.

This project was done with financial support from the Indo-Russian
International Long Term Programme of the Department of Science and
Technology, Government of India and the Russian Academy of Sciences.
Financial support in the initial phase from the Indian National Science
Academy exchange programme is also acknowledged. DVL acknowledges support
from the Joint Institute for VLBI in Europe for a visit there.
DG acknowledges support from the European Commission under TMR contract
No. ECBFMGECT950012. The VLA and VLBA are operated
by the National Radio Astronomy Observatory, a facility of the National
Science Foundation operated under cooperative agreement by Associated
Universities, Inc.  The European VLBI Network is a joint facility of
European, Chinese and other radio astronomy institutes funded by their
national research councils.  This research has made use of NASA's
Astrophysics Data System bibliographic services, the NASA/IPAC Extragalactic
Database (NED) which is operated by the Jet Propulsion Laboratory,
California Institute of Technology, under contract with NASA, and the SIMBAD
database, operated by CDS, Strasbourg, France.

\end{acknowledgements}

\bibliographystyle{aa}

\end{document}